\documentclass[%
 reprint,
 amsmath,amssymb,
 aps,
]{revtex4-2}
\usepackage{xcolor}
\usepackage{graphicx}
\usepackage{dcolumn}
\usepackage{bm}
\usepackage{verbatim}
\usepackage{xcolor, colortbl}
\usepackage{amsmath}
\usepackage{tcolorbox}
\usepackage{array}
\usepackage{xcolor,pict2e}
\usepackage{bbm}
\usepackage{bm}
\usepackage{dsfont}
\usepackage{graphicx}

\begin{document}


\title{Two dimensional effects  of laser interacting with  magnetized plasma}

 
%

\author{Laxman Prasad Goswami}
\email {goswami.laxman@gmail.com}
\affiliation{Department of Physics, Indian Institute of Technology Delhi, Hauz Khas, New Delhi 110016, India}

\author{Amita Das}
\email {amita@iitd.ac.in}
\affiliation{Department of Physics, Indian Institute of Technology Delhi, Hauz Khas, New Delhi 110016, India}

\author{Anuj Vijay}
\affiliation{Department of Physics, GLA University, Mathura 281406, India}
 





\begin{abstract}
Recent advancements in low-frequency short-pulse $CO_2$ lasers and the production of strong magnetic fields have made experimental studies on laser interactions with magnetized plasma a near-future possibility. Therefore, theoretical and numerical simulation studies have  been pursued lately in this direction [A. Das, Review of Modern Plasma Physics 4, 1 (2020)]
illustrating a host of novel phenomena related to laser energy absorption [Vashistha et al., New Journal of Physics, 22(6):063023 (2020); Goswami et al., Plasma Physics and Controlled Fusion 63, 115003 (2021)], harmonic generation 
[Maity et al., Journal of Plasma Physics, 87(5) (2021)], etc. However, most of these studies have been carried out in one-dimensional geometry with the laser having infinite transverse extent, and the plasma target was considered cold. This manuscript explores the manifestation of the 2-D and thermal effects on the problem of a laser interacting with magnetized plasma. As expected,  additional transverse ponderomotive force is shown to be operative.    A finite temperature of the target, along with  transverse density stratification generates, leads to diamagnetic drift for the two plasma species. The imbalance of this drift between the two species can be an additional effect leading to an enhancement of  laser energy absorption. The Particle  -  In - Cell (PIC) simulations with the OSIRIS4.0 platform is used to explore  these features.  


\end{abstract}

\maketitle

\section{Introduction}
\label{sec:Introduction}

The invention of high-power lasers \cite{mourou2014single, mourou2019nobel, nakajima2018seamless, strickland2019nobel} opens a broad spectrum of studies in electromagnetic (EM) wave interaction with matter \cite{kaw2017nonlinear, das2020laser, gong2019recent}. Laser plasma interaction is a rapidly growing field of research that has significant implications for numerous scientific and technological applications. The interaction between a high-intensity laser beam and a plasma can result in a variety of complex phenomena, including the generation of high-energy particles \cite{silva2004proton, macchi2013ion, macchi2013superintense}, the production of X-rays \cite{chen1993hot, yasuike2001hot}. It also has the potential for aiding laser fusion \cite{tanaka2000studies, kemp2014laser, campbell2017laser, clark2019three, craxton2015direct, hurricane2019approaching, kline2019progress, lindl1995development, merritt2019experimental, montgomery2016two, myatt2017wave, tabak1994ignition}. The problem of ignition in the context of laser fusion requires ion heating. The laser energy, however, gets transferred to the lighter electron species, and the ion heating occurs as a secondary process.  The collisional electron-ion interaction and/or interaction mediated by turbulent fields aids the process of the transfer of electron energy to ions. Both these processes are, however, of secondary nature wherein the laser energy is first transferred to electrons.  Lately, \cite{vashistha2020new}, it has been shown through simulations that electron motion can be restricted  with the help of strong magnetic fields. In that case,  one can then transfer laser energy directly to ions. The application of a strong magnetic field also enables low-frequency electromagnetic waves and/or lasers  to interact with the bulk plasma through the various permitted pass bands of the  magnetized dispersion relation.  The availability of low-frequency high-intensity pulsed $CO_2$ lasers and the possibility of generating Kilo Tesla order magnetic fields \cite{nakamura2018record} opens up the prospect of a new regime in which the electron's gyrofrequency can be higher than the laser frequency. This particular regime is good for direct laser energy transfer to ions. There are theoretical proposals \cite{korneev2015gigagauss, zosa2022100} indicating the possibility of generating  magnetic fields of the order of Mega Tesla. If this is achieved experimentally, even the heavier ions can get magnetized (their gyrofrequency could also be higher than the laser frequency). Such a regime has many other interesting fundamental prospects, such as complete transparency, which have been shown recently \cite{mandal2021transparency, goswami2021ponderomotive, mandal2021electromagnetic}. However, all these magnetized plasma studies have typically been restricted   only to the one-dimensional configuration. On the other hand, the 2-D effects in the context of unmagnetized plasma have been studied extensively, which  have shown complex interplay between the laser beam, the plasma, and the surrounding medium, leading to the formation of plasma channels \cite{ritchie1994relativistic}, self-focusing of the laser beam\cite{osman1999relativistic}, filamentation \cite{young1989filamentation}, surface instabilities \cite{liu2019high}. 

We endeavor to explore the possible 2-D features that may arise in magnetized plasma. 
The laser/EM wave is still considered incident normal to the target surface. Oblique incidence has not been considered in the present study. In particular, we have concentrated here on possible effects that may arise through the transverse variations that will be present. These may arise due to the finite focal spot size etc. For such a case, the transverse ponderomotive force will also be present due to the transverse laser intensity variations.  At very high intensity, the  transverse ponderomotive pressure might also lead to density stratification of the plasma along the transverse direction. This may also lead to diamagnetic drifts along the propagation direction in the 2-D geometry for a hot magnetized plasma. This drift might also lead  to charge separation, provided the two species have a different magnitude. The   electrostatic fluctuations thus generated would  lead  to  another possible absorption mechanism in the context of magnetized plasma. To illustrate the proof of principle concept  here, we have shown this feature by choosing a pre-specified density stratification profile and not the one arising through the transverse ponderomotive pressure of the EM wave, which could  be very weak to give a noticeable effect. 

The manuscript has been organized as follows. In section \ref{sec:Theory}, we have given the theoretical description of the two possible drifts where 2-D variations in the system are present  for a laser interacting with magnetized plasma.   Section \ref{sec:SimulationDetails} provides simulation details. We have discussed the simulation results for ponderomotive force and diamagnetic drifts in section \ref{sec:Ponderomotive} and \ref{sec:Diamagnetic}, respectively. Finally, we summarize and conclude in section \ref{sec:Conclusion}.

\section{Theoretical Description}
\label{sec:Theory}

\begin{figure*}
	\centering
	\includegraphics[width=6.0in]{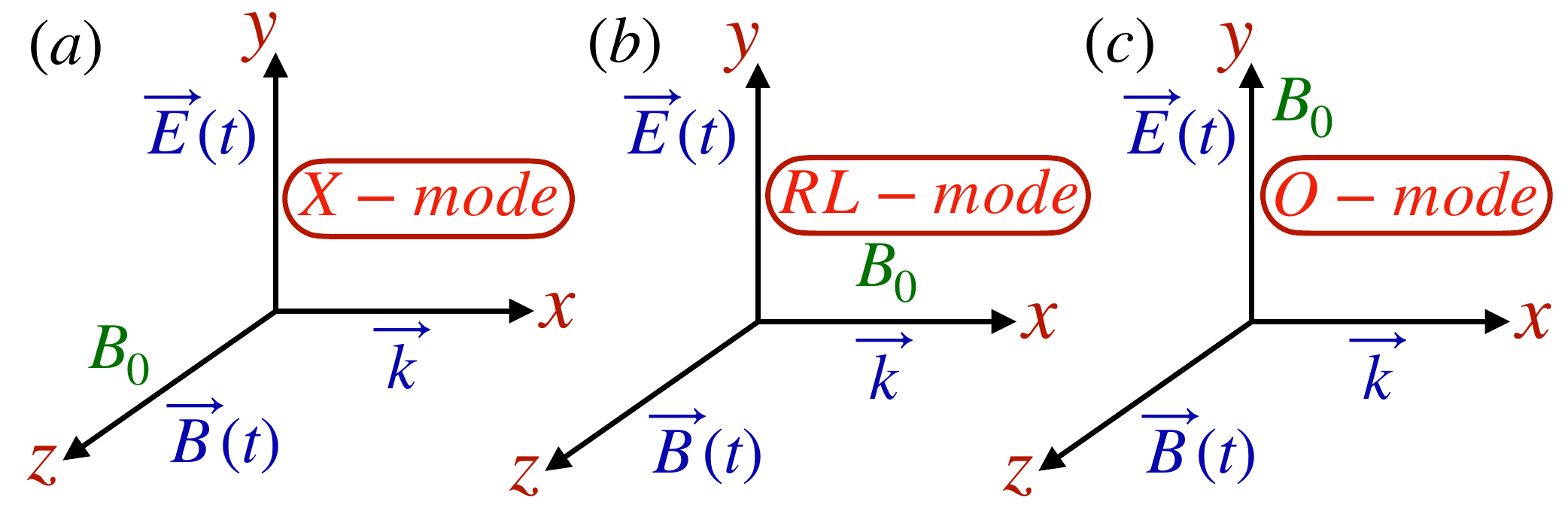}
	\caption{The figure shows the magnetized plasma geometry for (a) Extraordinary ($X-mode$) having external magnetic field ($B_0$) along the laser magnetic field direction ($\Vec{B}(t)$), (b) Right-Left Circularly polarized ($RL-mode$) where $B_0$ is acting along the laser propagation direction ($\Vec{k}$), and (c) Ordinary ($O-mode$) when $B_0$ is applied along the electric field ($\Vec{E}(t)$) of the incident laser pulse.}
	\label{fig:PlasmaGeometry}
\end{figure*}

\begin{figure*}
	\centering
	\includegraphics[width=6.0in]{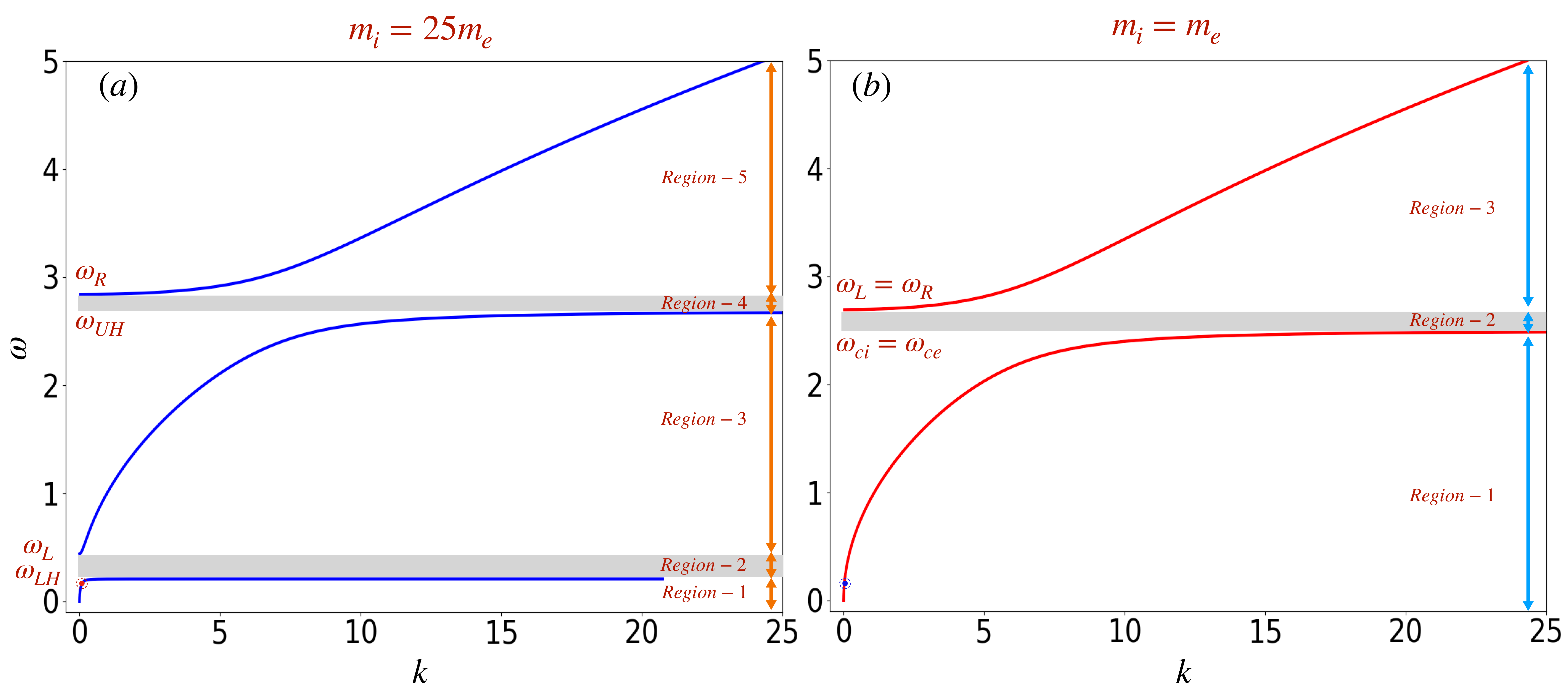}
	\caption{The figure shows the dispersion curve for X-mode both for (a) $m_i = 25m_e$ and (b) $m_i=m_e$, where $m_i$ and $m_e$ are the ion and electron masses. In this figure, $k$ and $\omega$ denote the wave number and frequency of the EM wave. $\omega_R$, $\omega_L$ are the right-hand and left-hand cutoffs. $\omega_{ce}$, $\omega_{ci}$ are the cyclotron frequencies of electron and ion. $\omega_{UH}$, $\omega_{LH}$ indicate the upper-hybrid and lower-hybrid resonance frequencies. Here the gray shading in regions-2 and 4 indicate the stop-band, while regions-1, 3, and 5 denote the passband. A red dot in each dispersion curve represents the laser frequency.}
	\label{fig:DispersionCurveXmode}
\end{figure*}

\begin{figure*}
	\centering
	\includegraphics[width=6.0in]{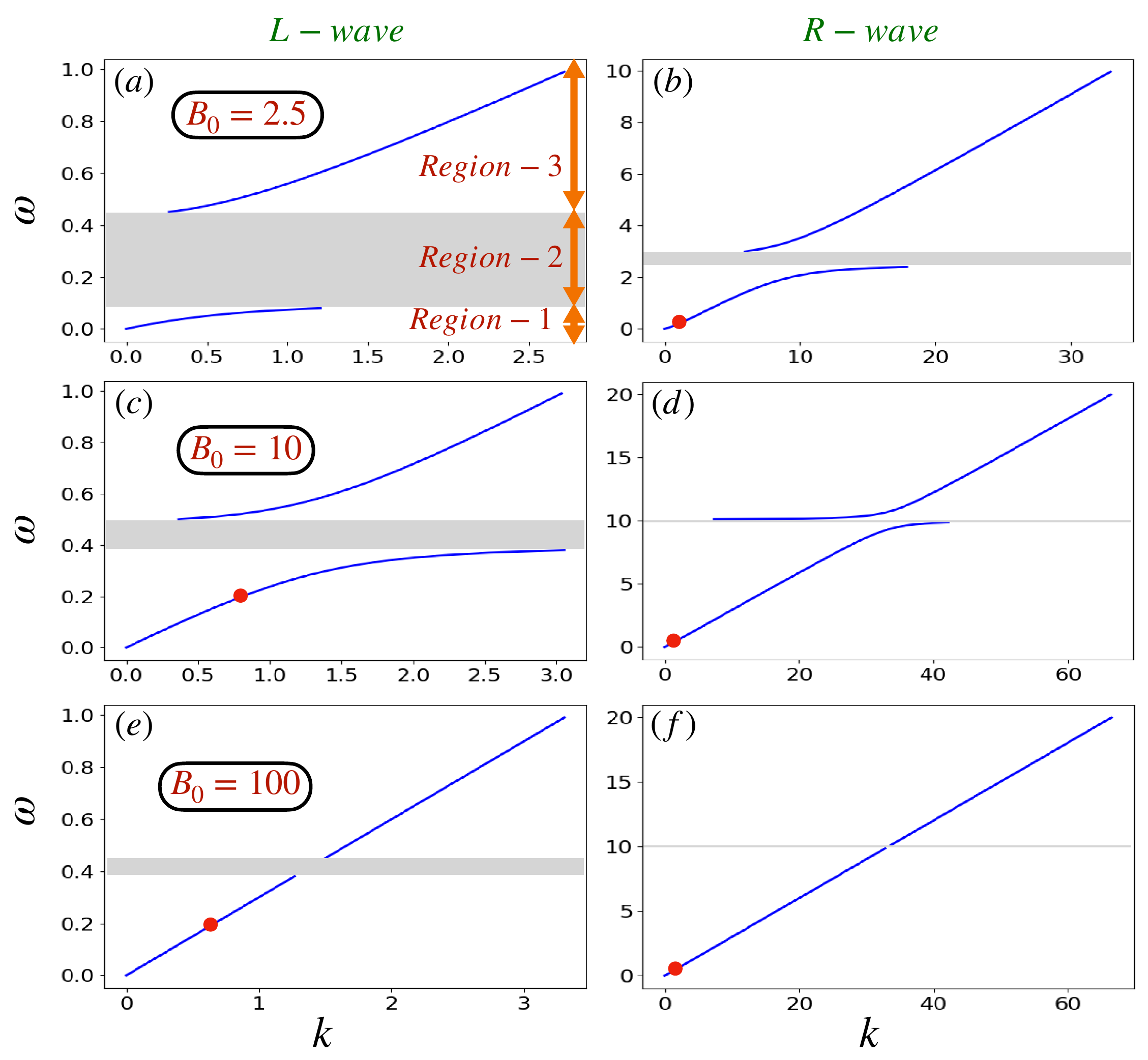}
	\caption{Figure shows the dispersion curve for $RL-mode$ geometry with applied external magnetic field $B_0 = 2.5$, $B_0 = 10$, and $B_0 = 100$. In this figure, $k$ and $\omega$ denote the wave number and frequency of the EM wave. Here the gray shading in regions-2 indicates the stop band, while regions-1 and 3 denote the passband. A red dot in each dispersion curve represents the laser frequency. The width of the stopband reduces with an increase in magnetic field strength.}
	\label{fig:RLDispersionCurves}
\end{figure*}

This section provides theoretical  details about a  magnetized plasma interacting with a laser. In particular, some possible effects which can only be observed in 2-dimensional simulations are explored in detail. There are three possible geometries (as shown in Fig.\ref{fig:PlasmaGeometry}) in which the external magnetic field can be oriented with respect to the propagation and electric field orientation of the laser field, and they are termed X-mode (extraordinary mode), RL-mode (right and left circularly polarised waves), and O-mode (ordinary wave) configurations. The EM wave satisfies a different dispersion relation for each of the orientations \cite{boyd2003physics}. The laser can penetrate an un-magnetized plasma only if its frequency is higher than the plasma frequency (i.e., underdense plasma with $\omega_l > \omega_{pe}$). For the overdense plasma (with $\omega_l < \omega_{pe}$), the laser can interact  only up to a width of skin depth. The application of magnetic field, however,  introduces certain pass-bands for example, Region-1 and 3 in the X-mode and RL-mode dispersion curves shown respectively in Fig.\ref{fig:DispersionCurveXmode} and Fig.\ref{fig:RLDispersionCurves}, that allows the propagation of EM wave even inside an overdense plasma medium. Thus, in this case, the laser can interact with the bulk of the plasma medium. By choosing the appropriate strength of the magnetic field,  one can selectively arrest the motion of electrons or elicit  a magnetized response from both  electron and  ion species. For instance, if laser frequency lies between the cyclotron frequency of electron and ion ($\omega_{ci}<\omega_l<\omega_{ce}$), electrons will show highly magnetized response, enabling ions to interact directly with the laser. For very high magnetic fields such that $\omega_{l}<\omega_{ci}<<\omega_{ce}$, both the species will show a magnetized response.  

\subsection{Electrostatic Wave Excitation }
We are essentially  seeking  the possibility of electrostatic waves/disturbances excitation by the laser in the plasma medium. The conversion/damping of such excitations  into  random kinetic energy of the  particles then leads to an irreversible transfer of laser energy into the plasma medium. The  mass and charge of the two plasma species (viz., electrons and ions) differ, which results in differences in their dynamics in the presence of electric and magnetic fields.   Recently, \cite{vashistha2020new} for X mode geometry, a  novel  method  of electrostatic wave generation was shown resulting from the disparate masses of the two species. The  difference between the $\vec{E}\times\vec{B_0}$ (where $B_0$ is the applied external magnetic field) drift between the two species due to  oscillating electric field of the laser leads to a finite current which is directed and also has spatial variations along the laser propagation direction. The spatial variation is associated with the finite wavelength of the oscillating electric field.   This leads to a finite divergence in current and hence charge separation. The electrostatic disturbance excites the plasma, which is crucial in laser energy absorption. 

There are other drifts and forces which can act differently on electrons and ions and could also result in the creation of charge separation, causing electrostatic disturbance. It should be noted that if the dynamics provide current in the transverse direction, then its divergence  is  associated with the transverse dimension of the laser focal spot size, considerably weaker than the longitudinal variations, which are sharper occurring at the scale of EM  wavelength. Some other possibilities leading to charge separation  have already been explored earlier. For instance, it has been shown that the ponderomotive force due to the  longitudinal spatial profile of the  laser also produces electrostatic disturbances\cite{goswami2021ponderomotive, goswami2022observations}.  The ponderomotive force acting on ions and electrons differs even in an un-magnetized plasma as it depends on the mass of the concerned species. Thus, even in an un-magnetized plasma, electrostatic disturbances are generated. However, the disturbances remain confined to the target surface in the overdense case, as the laser cannot penetrate the bulk region of the plasma. This force is operative in an external magnetic field over the bulk region. A detailed description of other possibilities leading to electrostatic disturbance generation in plasma under various conditions has been provided in Table-\ref{table:DriftApplicable}. The laser is incident normal to the target surface, and its spatial profile is finite only in one of the transverse directions, i.e., the one that lies in the plane of incidence. The applied external magnetic field is chosen to be homogeneous, and the plasma is supposed to have a finite constant temperature. The possibility of  density stratification along the longitudinal and the transverse direction has been considered. The shaded boxes in Table-\ref{table:DriftApplicable} show the configuration that can generate electrostatic disturbances only when 2-D variations in simulations are considered. Even the 1-D simulation with variations along the laser propagation direction suffices for other cases. 
In this work, we study these cases, which require two-dimensional considerations. The two possibilities are  transverse ponderomotive force and the diamagnetic drifts on which we now concentrate.  

The equation of motion of the two fluid systems of electrons and ions in the presence of a combined Electric ($\vec{E}$) and Magnetic ($\vec{B}$) field is given by:
\begin{equation}
   m_sn_s\left[\frac{\partial \vec{v}_s}{\partial t} + \left(\vec{v}_s\cdot\vec{\nabla}\right)\vec{v}_s\right] = q_sn_s\left(\vec{E} + \vec{v}_s\times\vec{B}\right) - \vec{\nabla}p_s.
    \label{Eq:momentum}
\end{equation}
The quantities $m_s$, $\vec{v}_s$, $n_s$, and $p_s$ represent plasma species' mass, velocity, density, and pressure. 
Where the suffix $s = e,i$ stands for the plasma electron and ion species, respectively. 

\subsection{Expression of Ponderomotive Force }
 In an   inhomogeneous oscillating electromagnetic field, the momentum equation Eq.(\ref{Eq:momentum}) can be simplified by averaging over high-frequency laser oscillations. The  averaging  of convective and $\vec{v} \times \vec{B}$ terms in Eq.(\ref{Eq:momentum}) leads to a  force (known as the ponderomotive force)  that  depends on the gradient of intensity and thus has  a nonlinear dependence of the field amplitude. The expression for ponderomotive force has been obtained by Goswami et al., \cite{goswami2021ponderomotive} in the absence of a pressure gradient term in the momentum equation Eq.(\ref{Eq:momentum}) for various orientations of the applied external magnetic field and the incident laser polarization.  Such a generalized expression  
 \cite{goswami2021ponderomotive} is given as:
\begin{equation}
    \frac{\partial\vec{v}_s}{\partial t} = -\frac{q_s^2 \nabla |E|^2}{2m_s^2 \beta\omega_l^2}\left[A_1 + \alpha^2A_2 - 2\alpha\frac{\omega_{cs}}{\omega_l} + \frac{\omega_{cs}^2}{\beta\omega_l^2}A_3\right],
\end{equation}
where, $A_1 = \left(1 - b_y^2\frac{\omega_{cs}^2}{\omega_l^2}\right)$, $A_2 = \left(1 - b_z^2\frac{\omega_{cs}^2}{\omega_l^2}\right)$, \\ $A_3 = \left(\alpha b_y - b_xb_y\frac{\omega_{cs}}{\omega}\right)^2 + \left(b_z - \alpha b_xb_z\frac{\omega_{cs}}{\omega}\right)^2$,
\begin{equation}
    \beta = 1 - \frac{\omega_{cs}^2}{\omega_l^2}\left(b_x^2 + b_y^2 + b_z^2\right).
\end{equation}
Here the applied  external magnetic field is  $\vec{B}_0 = b_x \hat{x} + b_y \hat{y} + b_z \hat{z}$, and  $\alpha$ takes the value of  $0$ for linear and $\pm 1$ for right and left circularly polarised EM wave. 
The $\nabla \mid E \mid^2$ represents the intensity gradient, which can be both along the propagation and the transverse direction. 
It is well known that based on the orientation of the applied magnetic field, magnetized plasma supports three different waves - X-mode, RL-mode, and O-mode. Fig.\ref{fig:PlasmaGeometry} clearly explains the three configurations. 
For X-mode configuration ($b_z = 1, b_x = b_y=0$), with linear polarisation ($\alpha=0$), ponderomotive acceleration takes the form:
\begin{equation}
    \frac{\partial\vec{v}_s}{\partial t} = -\frac{q_s^2 \nabla|E|^2}{2m_s^2\omega_l^2\left(1 - \frac{\omega_{cs}^2}{\omega_l^2}\right)^2}.
    \label{Eq:PondX}
\end{equation}

For RL-mode geometry ($b_x = 1, b_y = b_z=0$), with linear polarisation ($\alpha=0$), ponderomotive acceleration is given as:
\begin{equation}
    \frac{\partial\vec{v}_s}{\partial t} = -\frac{q_s^2 \nabla|E|^2}{2m_s^2\omega_l^2\left(1 - \frac{\omega_{cs}^2}{\omega_l^2}\right)}.
    \label{Eq:PondRL}
\end{equation}

\subsection{Diamagnetic Drift}
When the pressure gradient term is retained, we expect  diamagnetic drift to be present. We wish to explore the role of this particular term in the laser-plasma interaction process. The expression for the diamagnetic drift term in the presence of a time-independent magnetic field is well known. We obtain here a generalized expression of the same in the presence of both a constant applied magnetic field and the  oscillating magnetic field of the laser. 
Consider an inhomogeneous  plasma medium with density variations along $\hat{y}$. Electromagnetic radiation with   its oscillating  electric field and magnetic field as  $\vec{E} =  E_1\hat{y}$ and  $\vec{B} =  B_1\hat{z}$ respectively is incident on the medium. In addition,  an external constant magnetic field  $B_0 \hat{y}$ is also assumed to be present.  We separate the particle velocity as a sum of constant and oscillating parts, viz., $\vec{v} = \vec{v}_{0s} + \vec{v}_{1s}$. Where the subscript $0$ and $1$ respectively represent the constant and oscillatory parts of the velocity. Taking the cross product of the momentum equation Eq.\ref{Eq:momentum} with the magnetic field $\vec{B}$, we obtain:

\begin{eqnarray}
    m_sn_s\left[\frac{\partial \vec{v}_s}{\partial t} + \left(\vec{v}_s\cdot\vec{\nabla}\right)\vec{v}_s\right]\times\vec{B} &=& \nonumber \\  q_s n_s\left(\vec{E} + \vec{v}_s\times\vec{B}\right)\times\vec{B} 
   &-& \vec{\nabla}p_s\times\vec{B}.
    \label{Eq:momentumCrossB}
\end{eqnarray}
Consider zero order terms in Eq.(\ref{Eq:momentumCrossB}), we have
The zeroeth order stationary terms can be collected to have:  
\begin{equation}
     q_sn_s\left( \vec{v}_0\times\vec{B}_0\right)\times\vec{B}_0 - \vec{\nabla}p_{s0}\times\vec{B}_0 = 0,
\end{equation}
which leads to the well-known  diamagnetic fluid drift normal to the applied magnetic field as  \cite{chen1984introduction}
\begin{equation}
    \vec{v}_{0\perp s} =  - \frac{\vec{\nabla} p_{s0} \times \vec{B}_0}{q_sn_sB_0^2}.
    \label{Eq:constantBdrift}
\end{equation}
Assuming uniform temperature the pressure gradient is along the same direction as the density gradient which has been chosen to be along $\hat{y}$. Since the external magnetic field is also along $\hat{y}$, there is no drift in the absence of electromagnetic radiation, i.e.  $\vec{v}_{0\perp s} = 0$. Now balancing the first-order terms in the EM fields, we obtain from  Eq.\ref{Eq:momentumCrossB} as:
\begin{equation}
    m_sn_s\frac{\partial \vec{v}_{1s}}{\partial t}\times \vec{B}_0 = q_sn_s\left[\vec{E}_1\times\vec{B}_0 - \vec{v}_{1\perp s}B_0^2\right] - \vec{\nabla}p_{s0}\times\vec{B}_1.
    \label{Eq:firstOrder}
\end{equation}
Both $\vec{E_1}$ and $\vec{B_0}$ are directed along $\hat{y}$; hence the first term on the right-hand side  is zero. 
The X-component of Eq.\ref{Eq:firstOrder} gives:
\begin{equation}
    \frac{\partial {v}_{1zs}}{\partial t} = \frac{q_s}{m_s}{v}_{1xs}B_0 + \frac{1}{m_s n_s}\frac{B_1}{B_0} 
    \frac{\partial p_{s0}}{\partial y}.
    \label{Eq:Xcomp}
\end{equation}
Similarly, Z-component of Eq.\ref{Eq:firstOrder} on the other hand is:
\begin{equation}
    \frac{\partial {v}_{1xs}}{\partial t} = -\frac{q_s}{m_s}\left[{v}_{1zs}B_0\right].
  \label{Eq:Ycomp}
\end{equation}
Eliminating $v_{1zs}$, and replacing time derivatives by $-i \omega$ we get the expression for  $v_{1xs}$ as:
\begin{equation}
    v_{1xs} = \frac{\omega_{cs}^2}{\omega^2_{cs} - \omega^2}\left( \frac{1}{q_s n_s B_0^2}\right)\left[ \vec{\nabla} p_{s0} \times \vec{B}_1\right]_x.
    \label{Eq:MagnetizeDrift}
\end{equation}
It is clear that this drift is dependent on the sign of the charge.   The drift  has a $x$ dependence (through the oscillating $B_1$ field of the laser). Hence,   the current associated with this drift will have finite divergence. 
We thus expect this drift to cause charge separation in the plasma leading to the generation of electrostatic excitations.  The electrostatic wave couples to the irreversible transfer of  energy to particles through  wave breaking and phase mixing processes envisaged in the literature.  It is also important to note that since the drift is charge dependent, even when the charges have equal masses, (the case of electron-positron plasma for instance)  this drift will be equal and in opposite  directions for both the species  and will give rise to  a current. 
    \label{Eq:UnmagnetizeDiaMag}
The summary of all concerning  drifts and forces acting on the plasma species  for normal incidence of laser on the target, operative for 1-D  as well as the 2-D variations  have been provided in  Table-\ref{table:DriftApplicable}. 
\begin{table*}
	\caption{Applicability of difference in drifts for electrons and ions in different plasma configurations. In this table, the right tick ($\checkmark$) represents the presence of a drift, and the cross mark ($\times$) indicates that the drift is absent. s (or p) indicate the laser polarisation perpendicular (or parallel) to the plane of incidence. The longitudinal profile of the laser is considered Gaussian. External magnetic field $B_0$ is homogeneous throughout the simulation box. Plasma has a finite uniform temperature for the diamagnetic drift cases. The density gradient is considered both along the longitudinal and transverse directions. Gray rectangles indicate the plasma configurations where 2D simulations are required.}
	\label{table:DriftApplicable}
	\begin{center}
		\begin{tabular}{|c|c|c|c|c|c|c|c|c|}
    \hline
        \color{blue} Plasma & \color{blue} Transverse &  \color{blue} Laser & \multicolumn{6}{c|}{\color{blue}Charge separation along}\\
        \cline{4-9}
        \color{blue} Configuration & \color{blue} Laser & \color{blue}Polarisation &  \multicolumn{3}{c|}{\color{blue}Longitudinal Direction} & \multicolumn{3}{c|}{\color{blue}Transverse Direction}\\
        \cline{4-9}
		& \color{blue} Profile & & \color{blue}$\vec{E}\times\vec{B}_0$ &  \color{blue}Ponderomotive & \color{blue}Diamagnetic & \color{blue}$\vec{E}\times\vec{B}_0$ &  \color{blue}Ponderomotive & \color{blue}Diamagnetic\\	
	\hline
        &  & p & $\times$ & $\checkmark$ & \cellcolor{gray}$\checkmark$ &
        $\times$ & \cellcolor{gray}$\checkmark$ & \cellcolor{gray}$\checkmark$\\
        & Gaussian & s & $\times$ & $\checkmark$ & $\times$ &
        $\times$ & \cellcolor{gray}$\checkmark$ &$\times$\\
        \cline{2-9}
        \color{blue}Un-magnetize &  & p &$\times$ & $\checkmark$ & \cellcolor{gray}$\checkmark$ &
        $\times$ & $\times$ &\cellcolor{gray}$\checkmark$\\
        & Plane & s &$\times$ & $\checkmark$ & $\times$ &
        $\times$ & $\times$ &$\times$\\
    \hline
		&  & p & $\checkmark$ & $\checkmark$ & \cellcolor{gray}$\checkmark$ &
  $\times$ & \cellcolor{gray}$\checkmark$ & \cellcolor{gray}$\checkmark$\\
        & Gaussian & s & $\checkmark$ & $\checkmark$ & $\times$ &
        $\times$ & \cellcolor{gray}$\checkmark$ & $\times$\\
        \cline{2-9}
        \color{blue}X-mode &  & p & $\checkmark$ & $\checkmark$ & \cellcolor{gray}$\checkmark$ &
        $\times$ & $\times$ & \cellcolor{gray}$\checkmark$\\        
        & Plane & s & $\checkmark$ & $\checkmark$ & $\times$ &
        $\times$ & $\times$ & $\times$\\
    \hline
        &  & p & $\times$ & $\checkmark$ & \cellcolor{gray}$\checkmark$ &
        $\times$ & \cellcolor{gray}$\checkmark$ & \cellcolor{gray}$\checkmark$\\
        & Gaussian & s & $\times$ & $\checkmark$ & $\times$ &
        \cellcolor{gray}$\checkmark$ & \cellcolor{gray}$\checkmark$ & $\times$\\
        \cline{2-9}
        \color{blue}RL-mode &  & p & $\times$ & $\checkmark$ & \cellcolor{gray}$\checkmark$ &
        $\times$ & $\times$ & \cellcolor{gray}$\checkmark$ \\        
        & Plane & s & $\times$ & $\checkmark$ & $\times$ &
        \cellcolor{gray}$\checkmark$ & $\times$ & $\times$\\
	\hline
        &  & p & $\times$ & $\checkmark$ & \cellcolor{gray}$\checkmark$ &
        $\times$ & \cellcolor{gray}$\checkmark$ & \cellcolor{gray}$\checkmark$\\
        & Gaussian & s & $\times$ & $\checkmark$ & \cellcolor{gray}$\checkmark$ &
        $\times$ & \cellcolor{gray}$\checkmark$ & \cellcolor{gray}$\checkmark$\\
        \cline{2-9}
        \color{blue}O-mode &  & p & $\times$ & $\checkmark$ & \cellcolor{gray}$\checkmark$ &
        $\times$ & $\times$ & \cellcolor{gray}$\checkmark$ \\
        & Plane & s & $\times$ & $\checkmark$ & \cellcolor{gray}$\checkmark$ &
        $\times$ & $\times$ & \cellcolor{gray}$\checkmark$\\
	\hline
        &  & p & $\times$ & $\times$ & \cellcolor{gray}$\checkmark$ &
        $\times$ & $\times$ & \cellcolor{gray}$\checkmark$\\
        \color{blue}{All} & Gaussian & s & $\times$ & $\times$ & $\times$ &
        $\times$ & $\times$ & $\times$\\
        \cline{2-9}
        \color{blue}{($m_i=m_e$)} &  & p & $\times$ & $\times$ & \cellcolor{gray}$\checkmark$ &
        $\times$ & $\times$ & \cellcolor{gray}$\checkmark$ \\
        & Plane & s & $\times$ & $\times$ & $\times$ &
        $\times$ & $\times$ & $\times$\\
	\hline
		\end{tabular}	
	\end{center}
\end{table*}

\section{Simulation}
\label{sec:SimulationDetails}

We have employed the OSIRIS-4.0 framework \cite{hemker2000particle, fonseca2002osiris, fonseca2008one} for carrying out two-dimensional particle-in-cell (PIC) simulations. The PIC code OSIRIS uses normalized values of the various parameters. As a convention, length is normalized by skin depth ($c/\omega_{pe}$), frequency by electron plasma frequency ($\omega_{pe} = \sqrt{4\pi n_0 e^2/m_e}$), and the electric and magnetic fields by $m_ec\omega_{pe}e^{-1}$. Here $n_0$ and $c$, respectively, denote the plasma density and the velocity of the EM wave in a vacuum. For our studies, we have used parameters associated with a $CO_2$ laser pulse with a wavelength of $\lambda_l = 9.42 \mu m$ (that corresponds to laser frequency $\omega_l = 0.2\omega_{pe}$).  The laser profile is Gaussian, having a rise and fall time of $100\omega_{pe}^{-1}$ with the peak intensity of $I = 3.5\times 10^{15}Wcm^{-2}$ (corresponding to a relativistic factor $a_0 (= eE/m_e\omega_l c) < 1 (a_0=0.5)$). Some studies have also been carried out with varying laser intensity; however, it has always been chosen to be non-relativistic to avoid interference from  relativistic effects. Boundary conditions are chosen to be absorbing for  both longitudinal and transverse directions. The simulation parameters of plasma and laser in normalized units alongside their typical possible values in SI units are provided in Table-\ref{table:simulationtable}. 

\begin{table}
	\caption{Values of simulation parameters in normalized and SI units for typical laser parameters}
	\label{table:simulationtable}
	\begin{center}
		\begin{tabular}{|c|c|c|}
			\hline
			\color{red}Parameters		
			 & \color{red}Normalized & \color{red}SI unit \\
			\hline
			\hline	
			\multicolumn{3}{|c|}{\color{blue}Plasma Parameters} \\
			\hline
			\color{blue}$n_0$ & $1.0$ & $3\times10^{26} $ \\
			\hline
			\color{blue}$\omega_{pe}$ & $1.0$ & $10^{15}$ \\
			\hline
			\color{blue}$m_i$ & $25\ m_e$ & $2.27\times 10^{-29}$ \\
			\hline
			\color{blue}$\omega_{pi}$ & $0.2\ \omega_{pe}$ & $0.2\times10^{15} $\\
			\hline
			\multicolumn{3}{|c|}{\color{blue}Laser Parameters} \\
			\hline
			\color{blue}$\omega_{l}$ & $0.2\ \omega_{pe}$ & $0.2 \times10^{15} $\\
			\hline
			\color{blue}$\lambda_{l}$ & $31.4\ c/\omega_{pe}$ & $9.42\ \mu m$\\
			\hline
			\color{blue}Intensity &  $a_0 = 0.5$ &$3.5\times 10^{19}$\\
			\hline
		\end{tabular}	
	\end{center}
\end{table}

In these studies, we have followed the dynamics of both electrons and ions. The ion mass has been chosen to be  $25$ times the mass of electrons $m_i=25m_e$ to reduce computation time. For some investigations, we have also considered  electron-positron plasma for which $m_i=m_e$. The  well-known un-magnetized laser energy absorption schemes have been ruled out by carefully choosing the simulation geometry. Normally incident laser on a sharp plasma interface ensures the absence of vacuum and resonance heating schemes. By considering non-relativistic laser ($a_0<1$), the role of $\vec{J}\times\vec{B}$ electron heating is made negligible. 

In the following two sections, we study the physics associated with  transverse ponderomotive force and the diamagnetic drift on laser-plasma interaction. To observe these effects, 2-D simulations are required. 

\section{Transverse Ponderomotive Effects}
\label{sec:Ponderomotive}

\begin{figure*}
	\centering
	\includegraphics[width=6.0in]{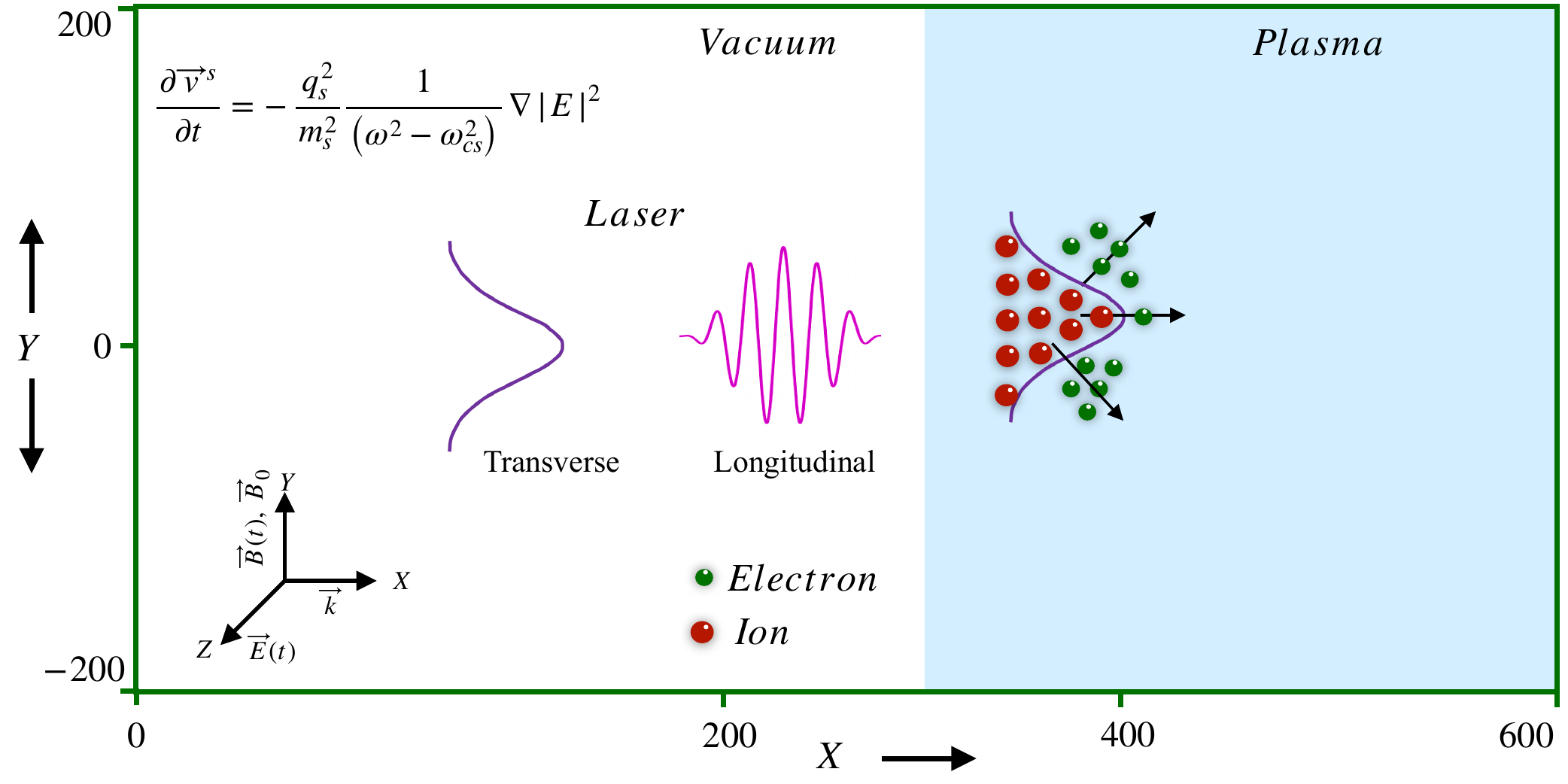}
	\caption{Schematics (not to scale) show the simulation setup and a summary of the physical process observed in our study. The laser propagating along the X-direction is incident from the left side of the simulation box on the vacuum-plasma interface at $x = 200$. We have chosen the X-mode geometry, where the external magnetic field $(B_0)$ is applied along the laser magnetic field direction $\Vec{B}(t)$. Green and red dots represent the plasma species - electrons, and ions. Initially, plasma is neutral, but charge separation is created in the transverse direction due to the transverse ponderomotive force of the incident laser pulse.}
	\label{fig:SchematicTransverse}
\end{figure*}

We describe the specifically chosen simulation geometry for studies pertaining to the role of transverse ponderomotive effects in  subsection (A). After that, in subsection (B) we discuss the observations and draw inferences. 

\subsection{Simulation Geometry for the study of transverse ponderomotive effect}
The schematic of the simulation geometry for ponderomotive force effects has been shown in Fig.\ref{fig:SchematicTransverse}. A rectangular simulation box in $X-Y$ plane of dimensions $L_x = 600$ and $L_y = 400$ with a grid size of $\Delta x = \Delta y = 0.1$ has been chosen to resolve the skin depth in our simulations. The number of particles per cell is taken to be $16$. There is a vacuum on the left side of the simulation box up to $x = 200$. There is plasma with uniform density between $x=200$ and $x=600$. For studies pertaining to this section, the external magnetic field $B_0$ has been chosen to be along $\hat{y}$. However, for some runs,  it has been chosen to be directed along the laser propagation direction $\hat{x}$ also. The laser pulse is incident normal to the plasma surface, and its electric field $\vec{E}$ is chosen to be directed along $\hat{z}$. The laser has a transverse profile along $\hat{y}$. Thus, the transverse ponderomotive force will be applicable along $\hat{y}$, leading to charge separation and a component of the electric field along $\hat{y}$, which can be easily distinguished from the laser electric field, which is 
chosen to be directed along $\hat{z}$. 
The plane and Gaussian transverse profile of laser intensity has been considered. This helps draw inferences about the role of transverse variation of laser intensity on observed absorption phenomena.  

\subsection{Observations and discussion}

\begin{figure*}
	\centering
	\includegraphics[width=6.0in]{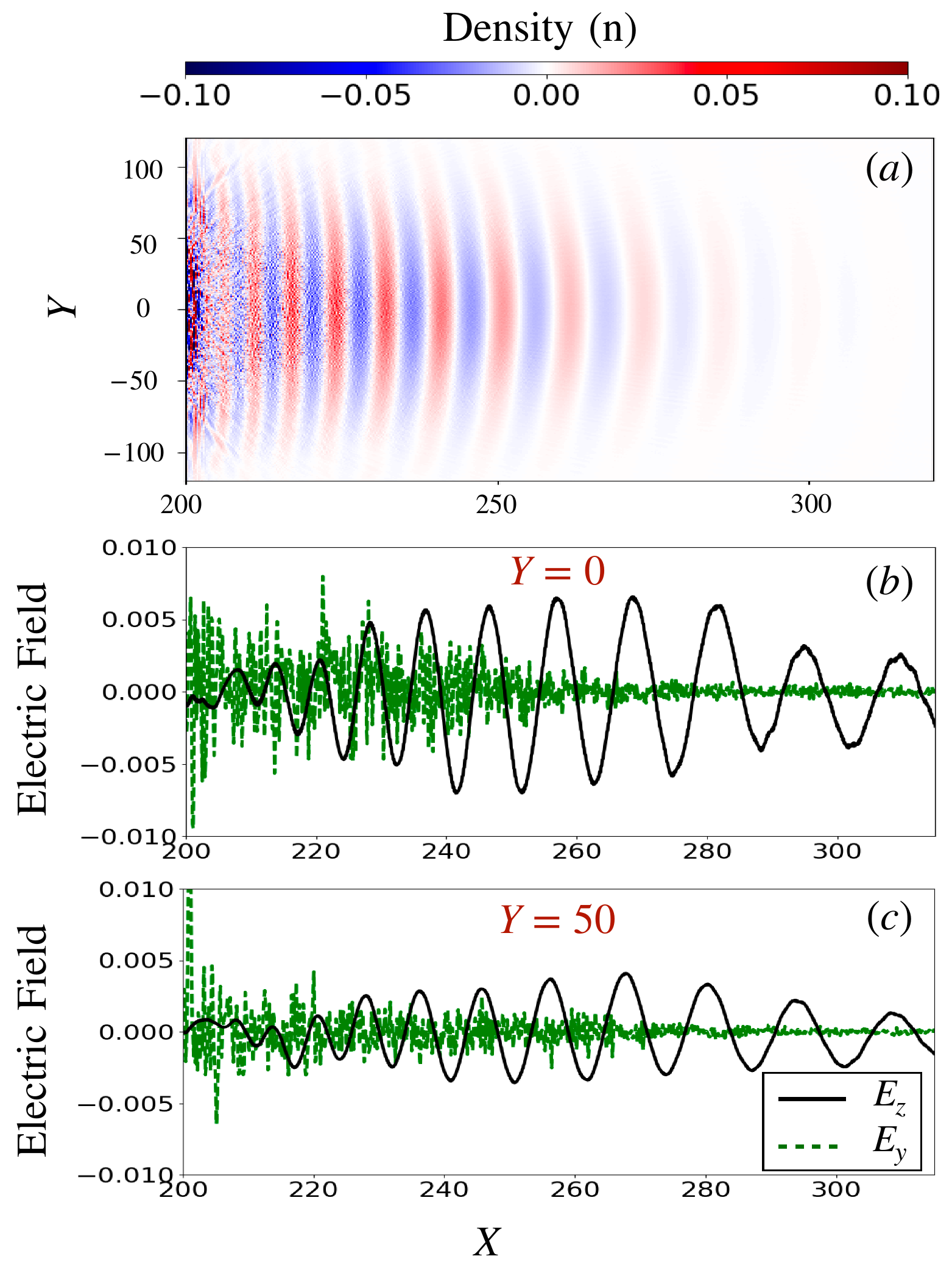}
	\caption{The figure shows (a) charge density (n) color plot with a color bar on the top and the line plots of electric field components $E_y$ (green dash line) and $E_z$ (solid black line) as a function of $X$ at (b) $Y=0$, and (c) $Y=50$ at time $t=600$ with $m_i = 25m_e$.}
	\label{fig:DensEyEz}
\end{figure*}

\begin{figure*}
	\centering
	\includegraphics[width=6.0in]{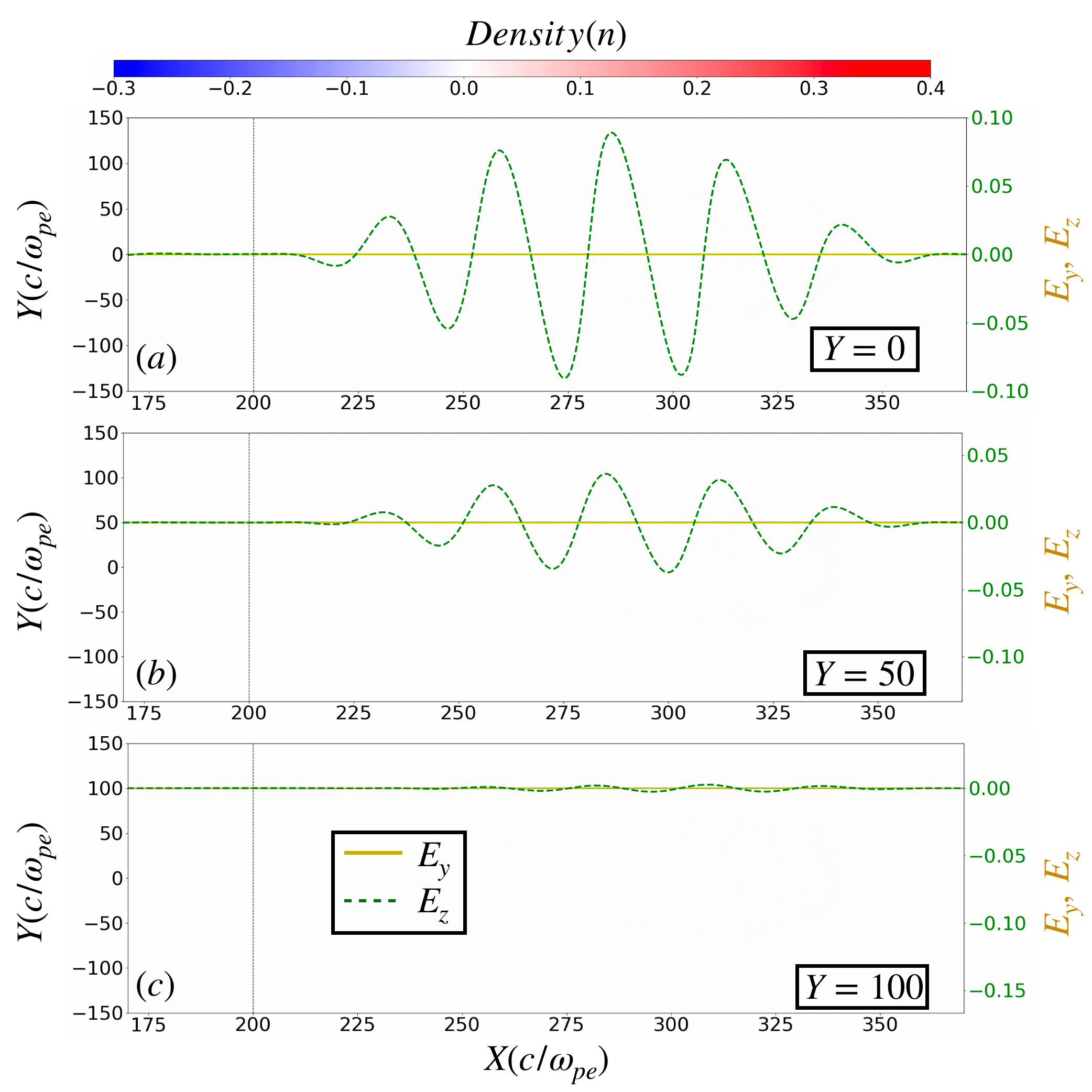}
	\caption{The figure shows the line plot of electric field components $E_y$ (solid cyan line) and $E_z$ (green dash line) as a function of $X$ superposed over the $2D$ color plot of charge density at time $t=200$ and (a) $Y=0$, (b) $Y=50$, and (c) $Y=100$ with $m_i = m_e$.}
	\label{fig:EPDensEyEz}
\end{figure*}

The spatial profile of charge density, and the $\hat{y}$ and $\hat{z}$ 
component of the electric field is shown in Fig.\ref{fig:DensEyEz} at time $t= 600$.  
The $\hat{y}$ component of the electric field is seen to get generated. The charge density has variations at the scale of laser wavelength along $\hat{x}$ and also has a transverse profile. 
In Fig.(\ref{fig:EPDensEyEz}), we have plotted a similar figure for the case for which $m_e = m_i$. For this case,  no charge density fluctuations get generated. Also, the electric field $E_y = 0$. This suggests that 
$E_y$ gets generated due to the ponderomotive force acting on the two charge species. When the two particle species have the same mass, this effect is absent, leading to no charge separation.   

\begin{figure*}
	\centering
	\includegraphics[width=6.0in]{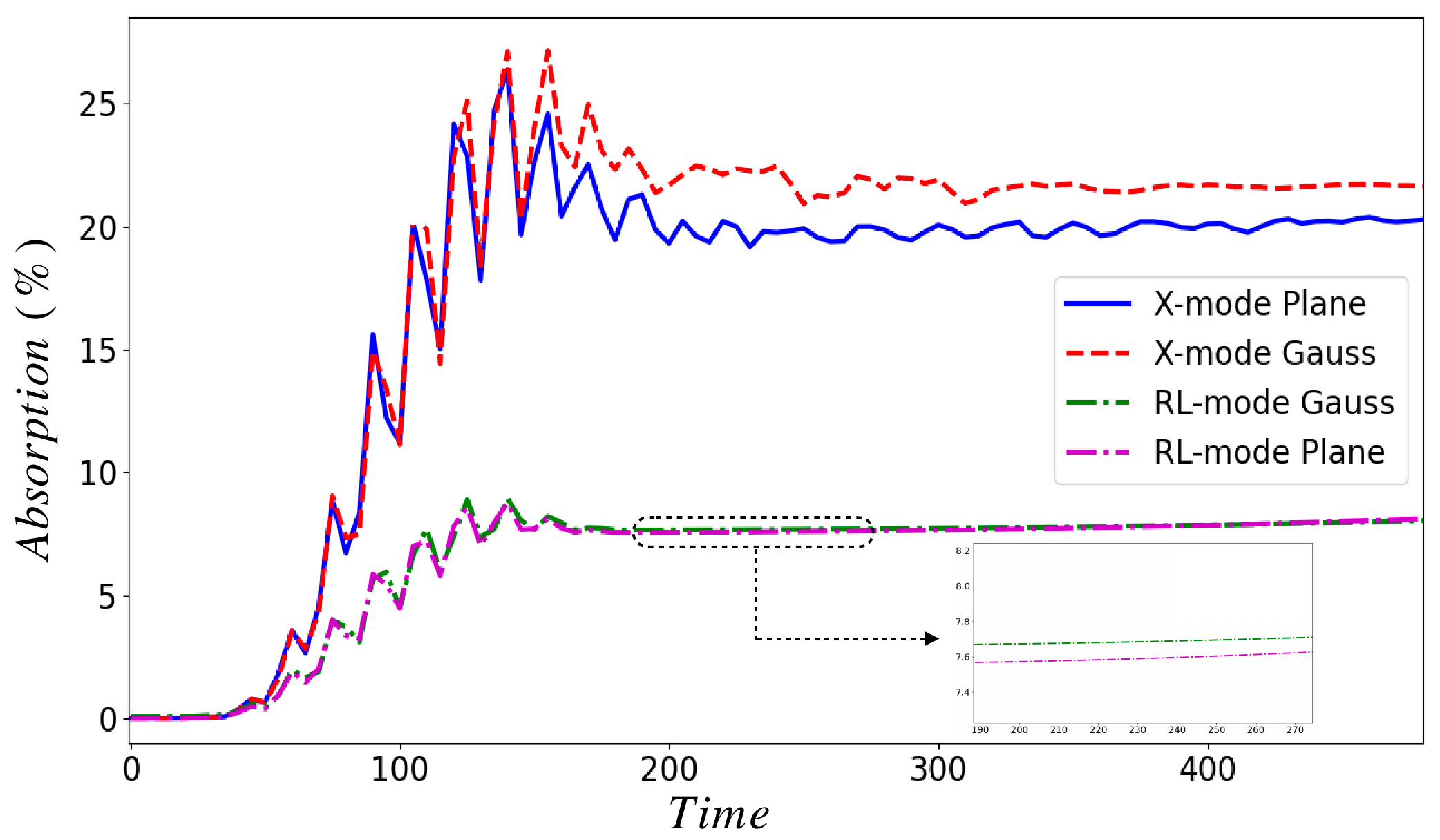}
	\caption{The figure shows the time evolution of percentage energy absorption in X-mode and RL-mode geometries with Gaussian and transverse plane profiles of the laser. It is evident from the figure that in both the geometries, the absorption is more with the Gaussian profile than with the Plane laser profile.}
	\label{fig:LaserProfileComparison}
\end{figure*}

In Fig.\ref{fig:LaserProfileComparison}, the percentage of energy absorption for both plane and Gaussian transverse profile of the laser has been shown. It can be observed that in the case having a transverse profile, the absorption is high. This is so even when the external magnetic field is chosen to be directed along the laser propagation direction. In this case, however, the difference between the absorption by plane and Gaussian profile is small,  and the difference is perceptible only from the zoomed inset plot of the same figure.  

\begin{figure*}
	\centering
	\includegraphics[width=6.5in]{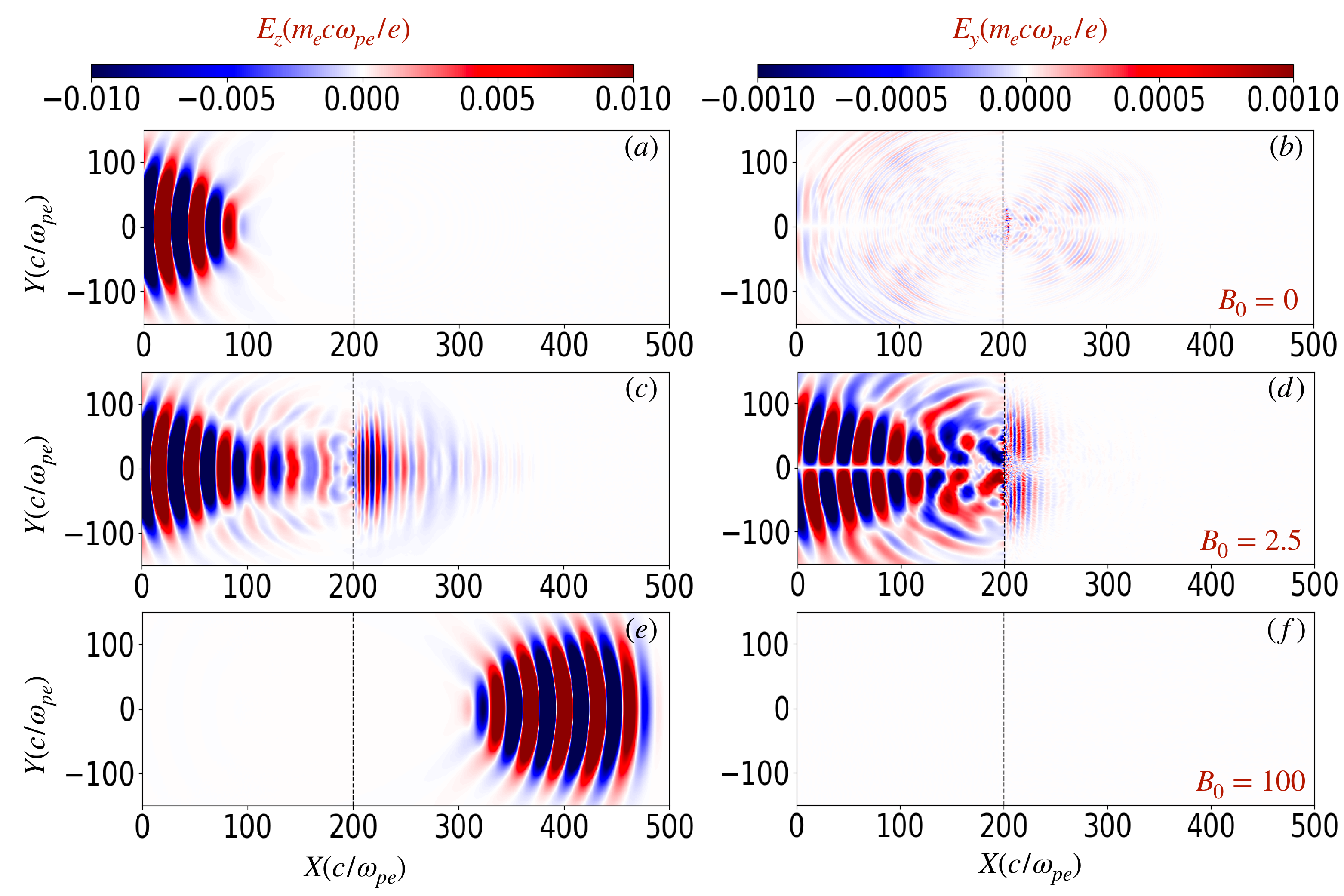}
	\caption{The figure shows the color plot of electric field components $E_z$ and $E_y$ at time $t=300$ with $m_i=25m_e$ for three different values of the external magnetic field $B_0=0,\ 2.5,\ 100$. It is evident from the figure that the excitation of the electric field $E_y$ component depends on the applied magnetic field. $E_y$ component is absent for very high magnetic fields, where the plasma shows transparency to the incident EM wave.}
	\label{fig:PondBComparision}
\end{figure*}

We have also carried out studies to understand the role of the  external magnetic field; we  consider magnetic field strength in the normalized units as $B_0=0,\ 2.5,\ 100$. Fig.\ref{fig:PondBComparision} gives the comparison of laser interaction with plasma for three different values of magnetic fields ($B_0 = 0$ subplot (a), (b), $B_0 = 2.5$ subplot (c), (d), and $B_0 = 100$ subplot (e), (f)). For the un-magnetized plasma $B_0 = 0$, the laser is completely reflected from the vacuum-plasma interface, and the electric field $E_y$ component is excited only at the surface of the plasma-vacuum boundary. When the external magnetic field is $B_0 = 2.5$, some part of the incident laser pulse is transmitted in the plasma, and this also excites the electric field $E_y$ component. For very large external magnetic fields $B_0=100$, one observes complete transmission of the pulse through the plasma. It is worth mentioning that this happens as a result of the shrinking of the stop band and the dispersion relation acquiring the vacuum form for the Electromagnetic waves. This effect has earlier been demonstrated by us in the context of 1-D simulation \cite{mandal2021electromagnetic, mandal2021transparency, goswami2021ponderomotive} for which the transverse extent of the laser was infinite. Our simulations here confirm that even in the 2-D simulations, it holds, and plasma behaves like a vacuum under strong magnetization.


\section{Diamagnetic Drift Effects}
\label{sec:Diamagnetic}

When finite temperature effects are incorporated for studying laser interacting with magnetized plasma, the diamagnetic drift will be present. 
This differential diamagnetic drift between the two species will generate current. If the current has a finite divergence, it can lead to electrostatic fluctuations aiding the absorption process. The effect of diamagnetic drift can be ascertained only in two and/or three dimensions. We carry out two-dimensional simulation studies here with a specific geometry described in the following subsection to illustrate the role of this particular drift in the absorption process. 

\subsection{Simulation Geometry}

\begin{figure*}
	\centering
	\includegraphics[width=6.0in]{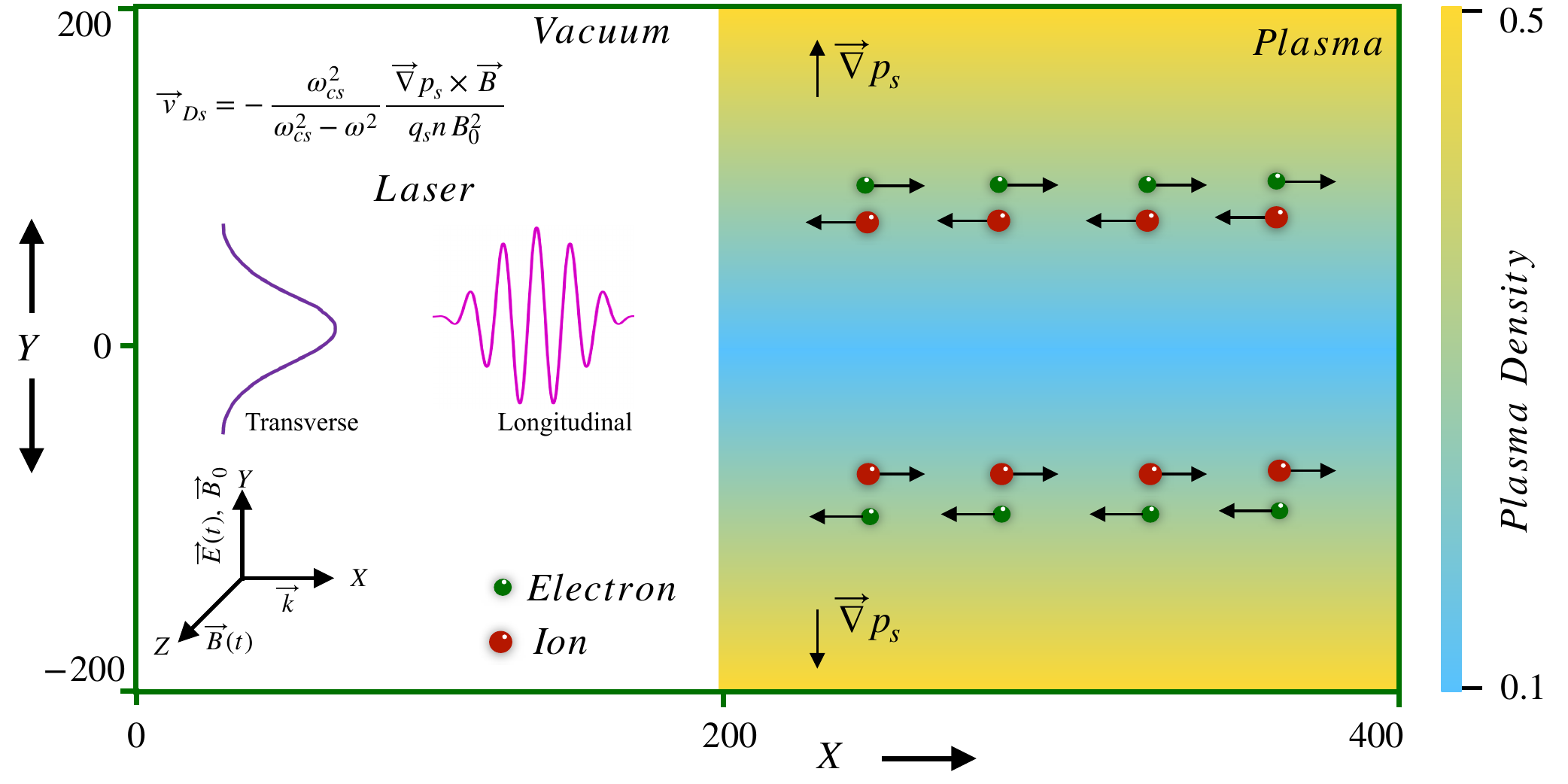}
	\caption{Schematics (not to scale) show the simulation setup for the diamagnetic drift studies. The color bar for plasma density is shown to the right. Green and red dots indicate the electron and ion species. A warm plasma is considered with an initial density gradient along $Y$. The laser propagating along the X-direction has a Gaussian profile in both longitudinal and transverse directions. We considered an O-mode (Ordinary) plasma configuration for this study with the external magnetic field applied along the laser field direction.}
	\label{fig:SchematicDiaMag}
\end{figure*}

The schematic of the simulation geometry for diamagnetic drift has been shown in Fig.\ref{fig:SchematicDiaMag}. Here a rectangular simulation box in $XY$ plane of dimensions $L_x = L_y = 400$  with a grid size of $\Delta x = \Delta y = 0.1$ has been chosen. This choice resolves the  skin depth adequately. The number of particles per cell is taken to be $8$. There is a vacuum on the left side of the simulation box up to $x = 200$. For  $ 200 < x < 400 $, plasma density is chosen to be finite and has a   functional  form defined by $n(y) = 1.5\times 10^{-3}|y|n_0$. Thus plasma density increases on both sides of the mid-plane of the box in the $y$ direction. The plasma is underdense up to the distance of $|y|=30$. The incident laser pulse is p-polarized with electric field vector oscillation along $\hat{y}$. We have chosen the O-mode geometry with the external applied along the laser electric field. This choice of simulation geometry may essentially provide proof of principle illustration of the diamagnetic drift-driven energy absorption mechanism  playing  a role in hot finite temperature magnetized plasmas. 


\subsection{Results and Discusion}



\begin{figure*}
	\centering
	\includegraphics[width=6.0in]{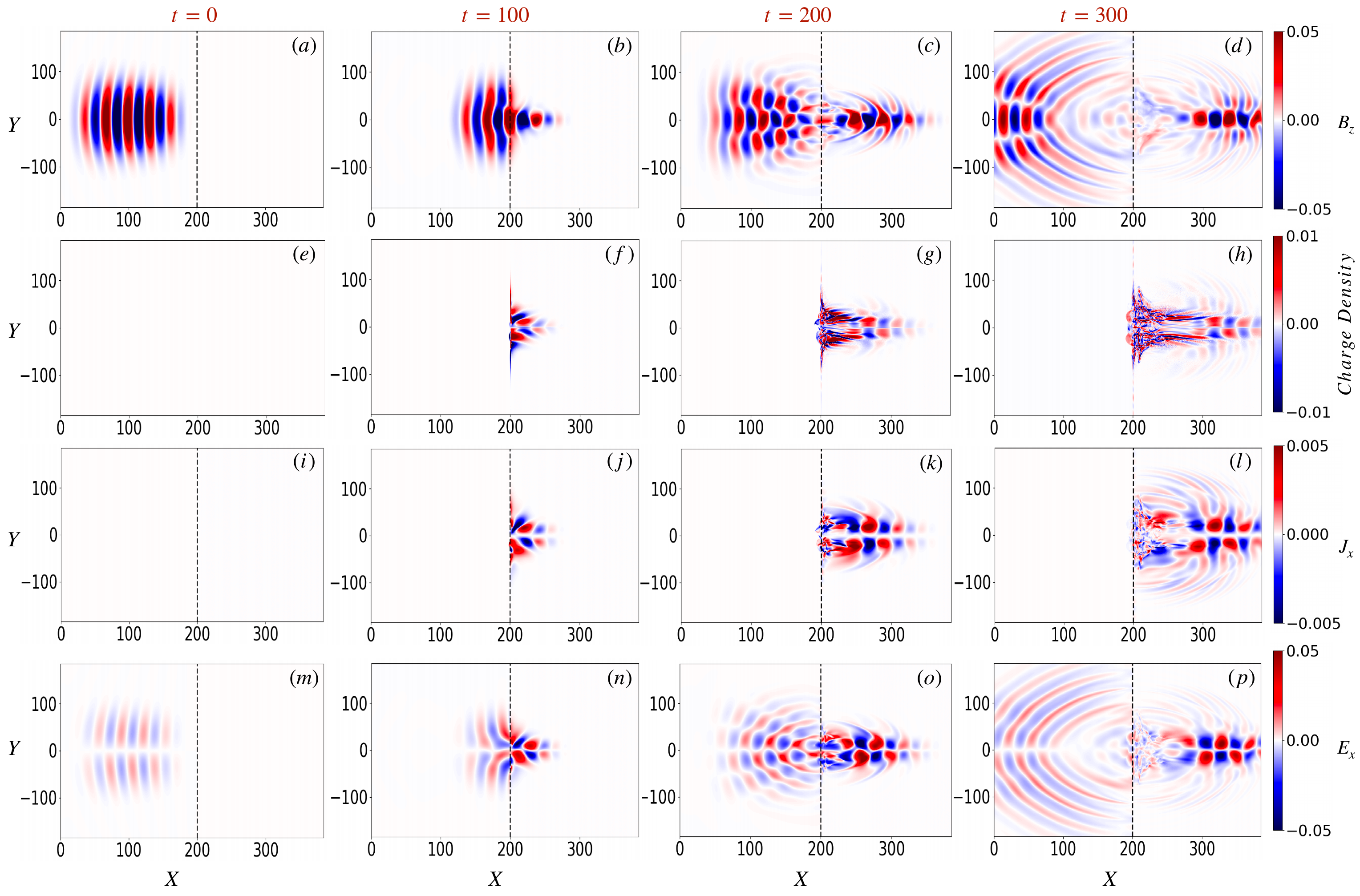}
	\caption{The figure shows the time evolution of magnetic field ($B_z$), charge density, current density ($J_x$), and the electric field ($E_x$) at time $t=0,\ 100,\ 200,\ 300$ for magnetized plasma (O-mode configuration) with $B_0=0.12$ (with Diamagnetic drift geometry). The color bar for each quantity is shown to the right of each figure. At the time $t=0$, the laser is in a vacuum region, and the charge and current density are zero in plasma. With laser interaction with plasma, diamagnetic drift effects lead to the excitation of charge density $n$, current density ${J_x}$, and electric field $E_x$ components.}
	\label{fig:diaOmode}
\end{figure*}

We have provided the time evolution of magnetic field $B_z$, charge density, current density $J_x$, and the electric field $E_x$ components in Fig.\ref{fig:diaOmode}. Before the interaction with a laser pulse  at time $t=0$, the charge and current density in the plasma are negligible. As the laser pulse interacts with the plasma, it generates charge and current density  (see subplots at time $t=100,\ 200$, and $300$). The excitation of charge and current density are associated with the generation of $E_x$ component of the electric field in the plasma. 


\begin{figure*}
	\centering
	\includegraphics[width=6.0in]{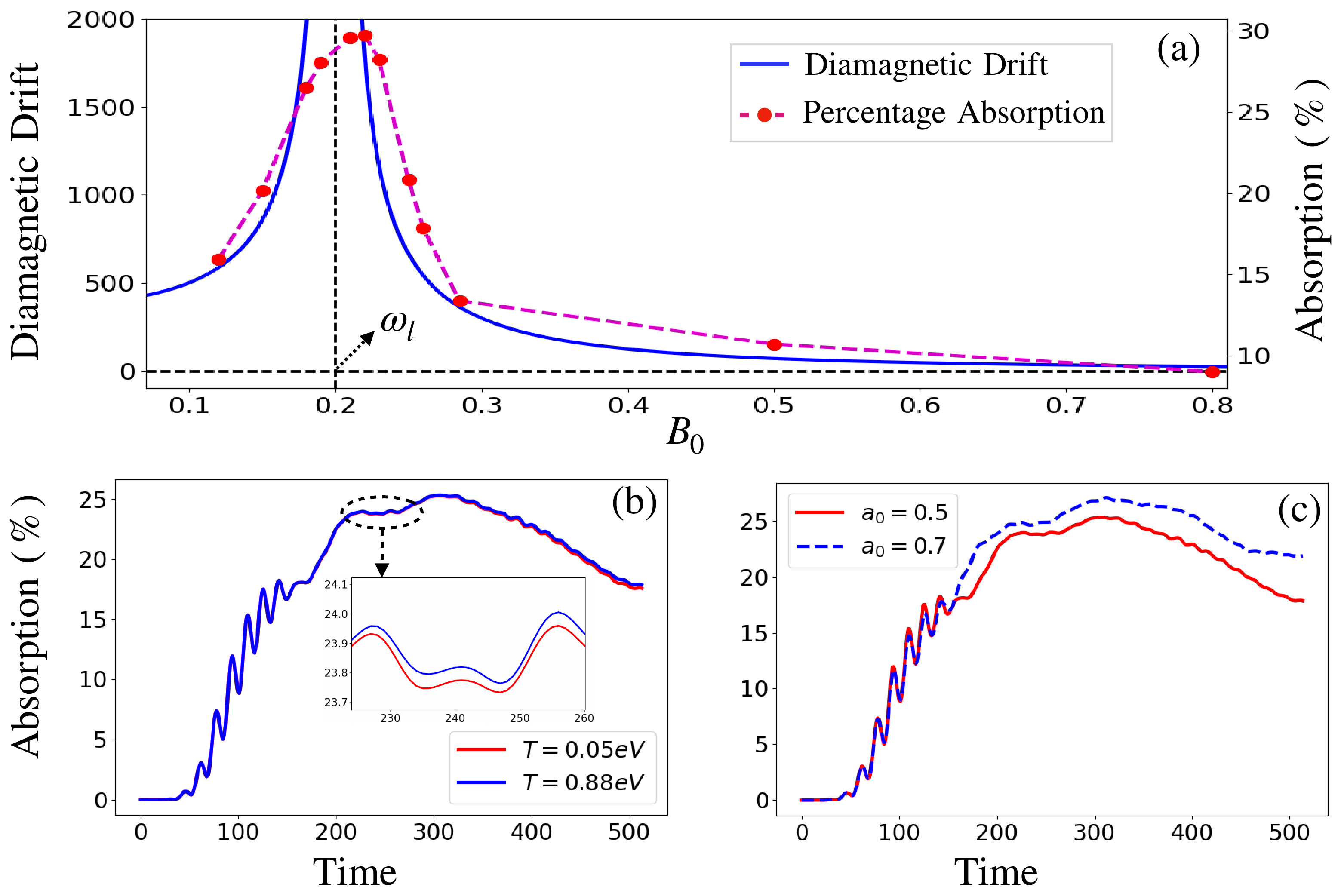}
	\caption{Subplot (a) of this figure shows the effects of the external magnetic field on the percentage energy absorption (red dot and magenta dash line). Solid blue lines in the same subplot indicate the analytic plot of diamagnetic drift. It is evident from the figure that when the cyclotron frequency of electrons is near to the laser frequency ($\omega_l$), the diamagnetic drift and hence the percentage energy absorption is more. The figure also shows the temporal evolution of percentage energy absorption with (b) different plasma temperatures (for $a_0 = 0.5,\ B_0=0.12$) and (c) different laser intensity (for $T=0.88eV,\ B_0=0.12$). It is evident from subplots (b) and (c) that the absorption is more at higher plasma temperatures and laser intensity.}
	\label{fig:DiaEnergyAbsorption}
\end{figure*}

We now discuss different physical ramifications arising from different masses and charges of different species. It is clear from Eq.\ref{Eq:MagnetizeDrift} that the diamagnetic drift for an oscillatory laser magnetic field differs for  ions and electrons due to their mass difference.  The difference in drift velocity, in turn, 
creates a finite current density in plasma. The current density is a function of space (due to laser profile) that leads to finite divergence of $\vec{J}$. 
The diamagnetic drift is directly proportional to the initial plasma temperature and the laser intensity. For this reason, $\vec{\nabla}\cdot\vec{J}$ will also be proportional to the plasma temperature (more for $T=0.88eV$ than $T=0.05eV$) and laser intensity (more for $a_0=0.7$ than $a_0=0.5$). The energy absorption is  directly linked to conversion to the electrostatic fluctuation of the laser energy. This effect is reflected in the percentage of laser energy absorption as shown in figures \ref{fig:DiaEnergyAbsorption}(b), \ref{fig:DiaEnergyAbsorption}(c). 

We further confirm the role of diamagnetic drift by plotting the expression of the diamagnetic drift as a function of the $B_0$ using Eq.(\ref{Eq:MagnetizeDrift}). On the same plot \ref{fig:DiaEnergyAbsorption}(a), we have also shown the percentage of laser energy absorption as a function of the magnetic field. The uncanny similarity in trend confirms the role of diamagnetic drift.

\begin{figure*}
	\centering
	\includegraphics[width=6.0in]{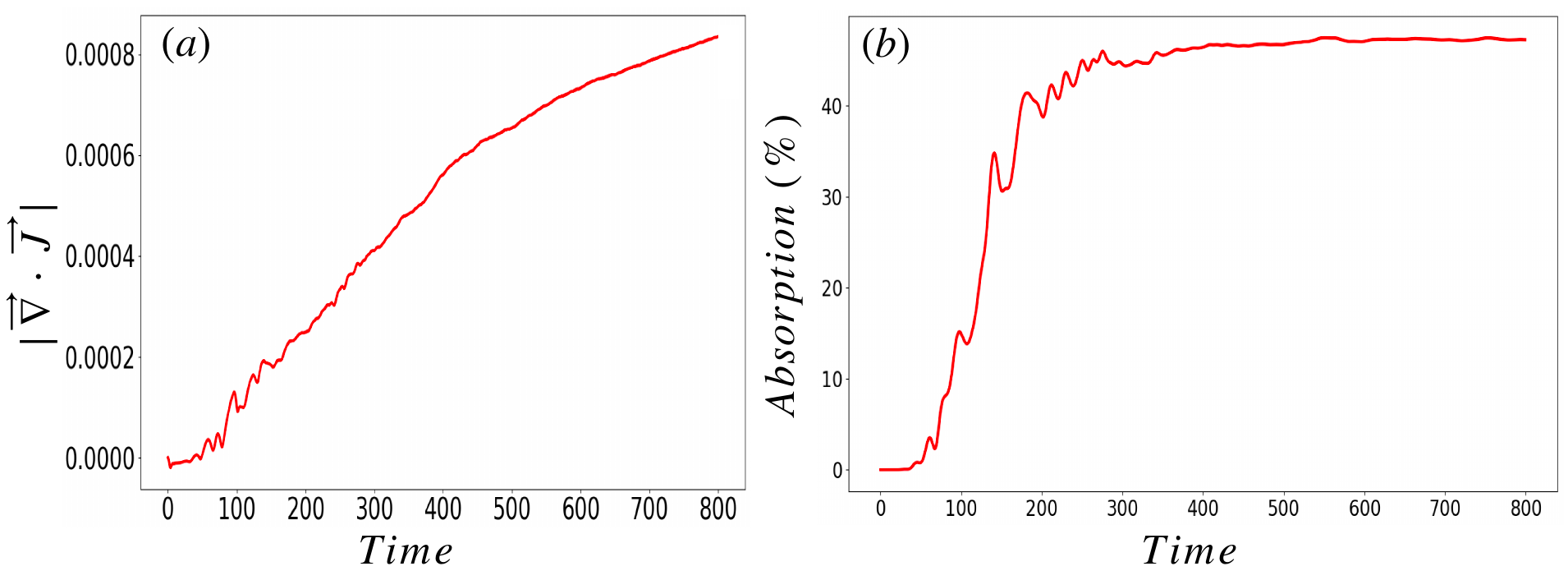}
	\caption{The figure shows the temporal evolution of (a) $|\vec{\nabla}\cdot\vec{J}|$ and (b) percentage energy absorption for electron-positron plasma ($m_i=m_e$). Here the initial plasma temperature is $T=0.88eV$, the laser intensity is $a_0=0.5$, and the external magnetic field is $B_0=0.12$. It is evident from the plots that though ponderomotive force-driven absorption is absent, diamagnetic effects are still present that provide effective laser energy absorption in plasma.}
	\label{fig:DiaEPplasma}
\end{figure*}

We have also carried out simulations with particles having equal mass and opposite and equal charges (e.g. like an electron-positron system).  For this case, the difference between the ponderomotive drift of the two species will be zero and will have no role to play.  The  diamagnetic drift-related charge separation, however, should still survive.  This is indeed found to be so, as simulations show finite   $\vec{\nabla}\cdot\vec{J}$ and  also laser energy  absorption also occurs as shown in Fig. (\ref{fig:DiaEPplasma}).    

\section{Conclusion}
\label{sec:Conclusion}

We have performed extensive two-dimensional PIC simulations to understand the role of transverse laser profile and temperature effects on laser energy absorption for a magnetized plasma. The generation of electrostatic waves/fields is known to aid the irreversible laser energy absorption process. The role of longitudinal ponderomotive pressure and the $\vec{E} \times \vec{B} $ drift has been investigated in some of the earlier studies \cite{vashistha2020new, goswami2021ponderomotive, goswami2022observations}. Here, in particular, we have identified the role of transverse ponderomotive force and the diamagnetic drift in generating electrostatic fields, thereby aiding the energy absorption process. 


\section*{Acknowledgements}

 The authors would like to acknowledge the OSIRIS Consortium, consisting of UCLA and IST (Lisbon, Portugal) for providing access to the OSIRIS4.0 framework which is the work supported by NSF ACI-1339893. This research work has been supported by the Core Research Grant No. CRG/2018/000624 of Department of Science and Technology (DST), Government of India. We also acknowledge support from the J C Bose Fellowship Grant of A D (JCB-000055/2017) from the Science and Engineering Research Board (SERB), Government of India. The authors thank IIT Delhi HPC facility for computational resources. Laxman thanks the Council for Scientific and Industrial Research (Grant no. 09/086(1442)/2020-EMR-I) for funding the research.


\begin{thebibliography}{41}%
\makeatletter
\providecommand \@ifxundefined [1]{%
 \@ifx{#1\undefined}
}%
\providecommand \@ifnum [1]{%
 \ifnum #1\expandafter \@firstoftwo
 \else \expandafter \@secondoftwo
 \fi
}%
\providecommand \@ifx [1]{%
 \ifx #1\expandafter \@firstoftwo
 \else \expandafter \@secondoftwo
 \fi
}%
\providecommand \natexlab [1]{#1}%
\providecommand \enquote  [1]{``#1''}%
\providecommand \bibnamefont  [1]{#1}%
\providecommand \bibfnamefont [1]{#1}%
\providecommand \citenamefont [1]{#1}%
\providecommand \href@noop [0]{\@secondoftwo}%
\providecommand \href [0]{\begingroup \@sanitize@url \@href}%
\providecommand \@href[1]{\@@startlink{#1}\@@href}%
\providecommand \@@href[1]{\endgroup#1\@@endlink}%
\providecommand \@sanitize@url [0]{\catcode `\\12\catcode `\$12\catcode
  `\&12\catcode `\#12\catcode `\^12\catcode `\_12\catcode `\%12\relax}%
\providecommand \@@startlink[1]{}%
\providecommand \@@endlink[0]{}%
\providecommand \url  [0]{\begingroup\@sanitize@url \@url }%
\providecommand \@url [1]{\endgroup\@href {#1}{\urlprefix }}%
\providecommand \urlprefix  [0]{URL }%
\providecommand \Eprint [0]{\href }%
\providecommand \doibase [0]{https://doi.org/}%
\providecommand \selectlanguage [0]{\@gobble}%
\providecommand \bibinfo  [0]{\@secondoftwo}%
\providecommand \bibfield  [0]{\@secondoftwo}%
\providecommand \translation [1]{[#1]}%
\providecommand \BibitemOpen [0]{}%
\providecommand \bibitemStop [0]{}%
\providecommand \bibitemNoStop [0]{.\EOS\space}%
\providecommand \EOS [0]{\spacefactor3000\relax}%
\providecommand \BibitemShut  [1]{\csname bibitem#1\endcsname}%
\let\auto@bib@innerbib\@empty
\bibitem [{\citenamefont {Mourou}\ \emph {et~al.}(2014)\citenamefont {Mourou},
  \citenamefont {Mironov}, \citenamefont {Khazanov},\ and\ \citenamefont
  {Sergeev}}]{mourou2014single}%
  \BibitemOpen
  \bibfield  {author} {\bibinfo {author} {\bibfnamefont {G.}~\bibnamefont
  {Mourou}}, \bibinfo {author} {\bibfnamefont {S.}~\bibnamefont {Mironov}},
  \bibinfo {author} {\bibfnamefont {E.}~\bibnamefont {Khazanov}},\ and\
  \bibinfo {author} {\bibfnamefont {A.}~\bibnamefont {Sergeev}},\ }\bibfield
  {title} {\bibinfo {title} {Single cycle thin film compressor opening the door
  to zeptosecond-exawatt physics},\ }\href@noop {} {\bibfield  {journal}
  {\bibinfo  {journal} {The European Physical Journal Special Topics}\ }\textbf
  {\bibinfo {volume} {223}},\ \bibinfo {pages} {1181} (\bibinfo {year}
  {2014})}\BibitemShut {NoStop}%
\bibitem [{\citenamefont {Mourou}(2019)}]{mourou2019nobel}%
  \BibitemOpen
  \bibfield  {author} {\bibinfo {author} {\bibfnamefont {G.}~\bibnamefont
  {Mourou}},\ }\bibfield  {title} {\bibinfo {title} {Nobel lecture: Extreme
  light physics and application},\ }\href@noop {} {\bibfield  {journal}
  {\bibinfo  {journal} {Reviews of Modern Physics}\ }\textbf {\bibinfo {volume}
  {91}},\ \bibinfo {pages} {030501} (\bibinfo {year} {2019})}\BibitemShut
  {NoStop}%
\bibitem [{\citenamefont {Nakajima}(2018)}]{nakajima2018seamless}%
  \BibitemOpen
  \bibfield  {author} {\bibinfo {author} {\bibfnamefont {K.}~\bibnamefont
  {Nakajima}},\ }\bibfield  {title} {\bibinfo {title} {Seamless multistage
  laser-plasma acceleration toward future high-energy colliders},\ }\href@noop
  {} {\bibfield  {journal} {\bibinfo  {journal} {Light, Science \&
  Applications}\ }\textbf {\bibinfo {volume} {7}} (\bibinfo {year}
  {2018})}\BibitemShut {NoStop}%
\bibitem [{\citenamefont {Strickland}(2019)}]{strickland2019nobel}%
  \BibitemOpen
  \bibfield  {author} {\bibinfo {author} {\bibfnamefont {D.}~\bibnamefont
  {Strickland}},\ }\bibfield  {title} {\bibinfo {title} {Nobel lecture:
  generating high-intensity ultrashort optical pulses},\ }\href@noop {}
  {\bibfield  {journal} {\bibinfo  {journal} {Reviews of Modern Physics}\
  }\textbf {\bibinfo {volume} {91}},\ \bibinfo {pages} {030502} (\bibinfo
  {year} {2019})}\BibitemShut {NoStop}%
\bibitem [{\citenamefont {Kaw}(2017)}]{kaw2017nonlinear}%
  \BibitemOpen
  \bibfield  {author} {\bibinfo {author} {\bibfnamefont {P.}~\bibnamefont
  {Kaw}},\ }\bibfield  {title} {\bibinfo {title} {Nonlinear laser--plasma
  interactions},\ }\href@noop {} {\bibfield  {journal} {\bibinfo  {journal}
  {Reviews of Modern Plasma Physics}\ }\textbf {\bibinfo {volume} {1}},\
  \bibinfo {pages} {1} (\bibinfo {year} {2017})}\BibitemShut {NoStop}%
\bibitem [{\citenamefont {Das}(2020)}]{das2020laser}%
  \BibitemOpen
  \bibfield  {author} {\bibinfo {author} {\bibfnamefont {A.}~\bibnamefont
  {Das}},\ }\bibfield  {title} {\bibinfo {title} {Laser plasma session:
  Aapps-dpp conference, 12--17 nov 2018, kanazawa},\ }\href@noop {} {\bibfield
  {journal} {\bibinfo  {journal} {Reviews of Modern Plasma Physics}\ }\textbf
  {\bibinfo {volume} {4}},\ \bibinfo {pages} {1} (\bibinfo {year}
  {2020})}\BibitemShut {NoStop}%
\bibitem [{\citenamefont {Gong}\ \emph {et~al.}(2019)\citenamefont {Gong},
  \citenamefont {Hao}, \citenamefont {Li}, \citenamefont {Yang}, \citenamefont
  {Li}, \citenamefont {Li}, \citenamefont {Guo}, \citenamefont {Zou},
  \citenamefont {Liu}, \citenamefont {Jiang} \emph {et~al.}}]{gong2019recent}%
  \BibitemOpen
  \bibfield  {author} {\bibinfo {author} {\bibfnamefont {T.}~\bibnamefont
  {Gong}}, \bibinfo {author} {\bibfnamefont {L.}~\bibnamefont {Hao}}, \bibinfo
  {author} {\bibfnamefont {Z.}~\bibnamefont {Li}}, \bibinfo {author}
  {\bibfnamefont {D.}~\bibnamefont {Yang}}, \bibinfo {author} {\bibfnamefont
  {S.}~\bibnamefont {Li}}, \bibinfo {author} {\bibfnamefont {X.}~\bibnamefont
  {Li}}, \bibinfo {author} {\bibfnamefont {L.}~\bibnamefont {Guo}}, \bibinfo
  {author} {\bibfnamefont {S.}~\bibnamefont {Zou}}, \bibinfo {author}
  {\bibfnamefont {Y.}~\bibnamefont {Liu}}, \bibinfo {author} {\bibfnamefont
  {X.}~\bibnamefont {Jiang}}, \emph {et~al.},\ }\bibfield  {title} {\bibinfo
  {title} {Recent research progress of laser plasma interactions in shenguang
  laser facilities},\ }\href@noop {} {\bibfield  {journal} {\bibinfo  {journal}
  {Matter and Radiation at Extremes}\ }\textbf {\bibinfo {volume} {4}},\
  \bibinfo {pages} {055202} (\bibinfo {year} {2019})}\BibitemShut {NoStop}%
\bibitem [{\citenamefont {Silva}\ \emph {et~al.}(2004)\citenamefont {Silva},
  \citenamefont {Marti}, \citenamefont {Davies}, \citenamefont {Fonseca},
  \citenamefont {Ren}, \citenamefont {Tsung},\ and\ \citenamefont
  {Mori}}]{silva2004proton}%
  \BibitemOpen
  \bibfield  {author} {\bibinfo {author} {\bibfnamefont {L.~O.}\ \bibnamefont
  {Silva}}, \bibinfo {author} {\bibfnamefont {M.}~\bibnamefont {Marti}},
  \bibinfo {author} {\bibfnamefont {J.~R.}\ \bibnamefont {Davies}}, \bibinfo
  {author} {\bibfnamefont {R.~A.}\ \bibnamefont {Fonseca}}, \bibinfo {author}
  {\bibfnamefont {C.}~\bibnamefont {Ren}}, \bibinfo {author} {\bibfnamefont
  {F.~S.}\ \bibnamefont {Tsung}},\ and\ \bibinfo {author} {\bibfnamefont
  {W.~B.}\ \bibnamefont {Mori}},\ }\bibfield  {title} {\bibinfo {title} {Proton
  shock acceleration in laser-plasma interactions},\ }\href@noop {} {\bibfield
  {journal} {\bibinfo  {journal} {Physical Review Letters}\ }\textbf {\bibinfo
  {volume} {92}},\ \bibinfo {pages} {015002} (\bibinfo {year}
  {2004})}\BibitemShut {NoStop}%
\bibitem [{\citenamefont {Macchi}\ \emph {et~al.}(2013)\citenamefont {Macchi},
  \citenamefont {Borghesi},\ and\ \citenamefont {Passoni}}]{macchi2013ion}%
  \BibitemOpen
  \bibfield  {author} {\bibinfo {author} {\bibfnamefont {A.}~\bibnamefont
  {Macchi}}, \bibinfo {author} {\bibfnamefont {M.}~\bibnamefont {Borghesi}},\
  and\ \bibinfo {author} {\bibfnamefont {M.}~\bibnamefont {Passoni}},\
  }\bibfield  {title} {\bibinfo {title} {Ion acceleration by superintense
  laser-plasma interaction},\ }\href@noop {} {\bibfield  {journal} {\bibinfo
  {journal} {Reviews of Modern Physics}\ }\textbf {\bibinfo {volume} {85}},\
  \bibinfo {pages} {751} (\bibinfo {year} {2013})}\BibitemShut {NoStop}%
\bibitem [{\citenamefont {Macchi}(2013)}]{macchi2013superintense}%
  \BibitemOpen
  \bibfield  {author} {\bibinfo {author} {\bibfnamefont {A.}~\bibnamefont
  {Macchi}},\ }\href@noop {} {\emph {\bibinfo {title} {A superintense
  laser-plasma interaction theory primer}}}\ (\bibinfo  {publisher} {Springer
  Science \& Business Media},\ \bibinfo {year} {2013})\BibitemShut {NoStop}%
\bibitem [{\citenamefont {Chen}\ \emph {et~al.}(1993)\citenamefont {Chen},
  \citenamefont {Soom}, \citenamefont {Yaakobi}, \citenamefont {Uchida},\ and\
  \citenamefont {Meyerhofer}}]{chen1993hot}%
  \BibitemOpen
  \bibfield  {author} {\bibinfo {author} {\bibfnamefont {H.}~\bibnamefont
  {Chen}}, \bibinfo {author} {\bibfnamefont {B.}~\bibnamefont {Soom}}, \bibinfo
  {author} {\bibfnamefont {B.}~\bibnamefont {Yaakobi}}, \bibinfo {author}
  {\bibfnamefont {S.}~\bibnamefont {Uchida}},\ and\ \bibinfo {author}
  {\bibfnamefont {D.}~\bibnamefont {Meyerhofer}},\ }\bibfield  {title}
  {\bibinfo {title} {Hot-electron characterization from k$\alpha$ measurements
  in high-contrast, p-polarized, picosecond laser-plasma interactions},\
  }\href@noop {} {\bibfield  {journal} {\bibinfo  {journal} {Physical review
  letters}\ }\textbf {\bibinfo {volume} {70}},\ \bibinfo {pages} {3431}
  (\bibinfo {year} {1993})}\BibitemShut {NoStop}%
\bibitem [{\citenamefont {Yasuike}\ \emph {et~al.}(2001)\citenamefont
  {Yasuike}, \citenamefont {Key}, \citenamefont {Hatchett}, \citenamefont
  {Snavely},\ and\ \citenamefont {Wharton}}]{yasuike2001hot}%
  \BibitemOpen
  \bibfield  {author} {\bibinfo {author} {\bibfnamefont {K.}~\bibnamefont
  {Yasuike}}, \bibinfo {author} {\bibfnamefont {M.}~\bibnamefont {Key}},
  \bibinfo {author} {\bibfnamefont {S.}~\bibnamefont {Hatchett}}, \bibinfo
  {author} {\bibfnamefont {R.}~\bibnamefont {Snavely}},\ and\ \bibinfo {author}
  {\bibfnamefont {K.}~\bibnamefont {Wharton}},\ }\bibfield  {title} {\bibinfo
  {title} {Hot electron diagnostic in a solid laser target by k-shell lines
  measurement from ultraintense laser--plasma interactions (3$\times$ 10 20  w/cm 2,$\leq$ 400 j)},\ }\href@noop {} {\bibfield  {journal} {\bibinfo
  {journal} {Review of Scientific Instruments}\ }\textbf {\bibinfo {volume}
  {72}},\ \bibinfo {pages} {1236} (\bibinfo {year} {2001})}\BibitemShut
  {NoStop}%
\bibitem [{\citenamefont {Tanaka}\ \emph {et~al.}(2000)\citenamefont {Tanaka},
  \citenamefont {Kodama}, \citenamefont {Fujita}, \citenamefont {Heya},
  \citenamefont {Izumi}, \citenamefont {Kato}, \citenamefont {Kitagawa},
  \citenamefont {Mima}, \citenamefont {Miyanaga}, \citenamefont {Norimatsu}
  \emph {et~al.}}]{tanaka2000studies}%
  \BibitemOpen
  \bibfield  {author} {\bibinfo {author} {\bibfnamefont {K.}~\bibnamefont
  {Tanaka}}, \bibinfo {author} {\bibfnamefont {R.}~\bibnamefont {Kodama}},
  \bibinfo {author} {\bibfnamefont {H.}~\bibnamefont {Fujita}}, \bibinfo
  {author} {\bibfnamefont {M.}~\bibnamefont {Heya}}, \bibinfo {author}
  {\bibfnamefont {N.}~\bibnamefont {Izumi}}, \bibinfo {author} {\bibfnamefont
  {Y.}~\bibnamefont {Kato}}, \bibinfo {author} {\bibfnamefont {Y.}~\bibnamefont
  {Kitagawa}}, \bibinfo {author} {\bibfnamefont {K.}~\bibnamefont {Mima}},
  \bibinfo {author} {\bibfnamefont {N.}~\bibnamefont {Miyanaga}}, \bibinfo
  {author} {\bibfnamefont {T.}~\bibnamefont {Norimatsu}}, \emph {et~al.},\
  }\bibfield  {title} {\bibinfo {title} {Studies of ultra-intense laser plasma
  interactions for fast ignition},\ }\href@noop {} {\bibfield  {journal}
  {\bibinfo  {journal} {Physics of Plasmas}\ }\textbf {\bibinfo {volume} {7}},\
  \bibinfo {pages} {2014} (\bibinfo {year} {2000})}\BibitemShut {NoStop}%
\bibitem [{\citenamefont {Kemp}\ \emph {et~al.}(2014)\citenamefont {Kemp},
  \citenamefont {Fiuza}, \citenamefont {Debayle}, \citenamefont {Johzaki},
  \citenamefont {Mori}, \citenamefont {Patel}, \citenamefont {Sentoku},\ and\
  \citenamefont {Silva}}]{kemp2014laser}%
  \BibitemOpen
  \bibfield  {author} {\bibinfo {author} {\bibfnamefont {A.}~\bibnamefont
  {Kemp}}, \bibinfo {author} {\bibfnamefont {F.}~\bibnamefont {Fiuza}},
  \bibinfo {author} {\bibfnamefont {A.}~\bibnamefont {Debayle}}, \bibinfo
  {author} {\bibfnamefont {T.}~\bibnamefont {Johzaki}}, \bibinfo {author}
  {\bibfnamefont {W.}~\bibnamefont {Mori}}, \bibinfo {author} {\bibfnamefont
  {P.}~\bibnamefont {Patel}}, \bibinfo {author} {\bibfnamefont
  {Y.}~\bibnamefont {Sentoku}},\ and\ \bibinfo {author} {\bibfnamefont
  {L.}~\bibnamefont {Silva}},\ }\bibfield  {title} {\bibinfo {title}
  {Laser--plasma interactions for fast ignition},\ }\href@noop {} {\bibfield
  {journal} {\bibinfo  {journal} {Nuclear Fusion}\ }\textbf {\bibinfo {volume}
  {54}},\ \bibinfo {pages} {054002} (\bibinfo {year} {2014})}\BibitemShut
  {NoStop}%
\bibitem [{\citenamefont {Campbell}\ \emph {et~al.}(2017)\citenamefont
  {Campbell}, \citenamefont {Goncharov}, \citenamefont {Sangster},
  \citenamefont {Regan}, \citenamefont {Radha}, \citenamefont {Betti},
  \citenamefont {Myatt}, \citenamefont {Froula}, \citenamefont {Rosenberg},
  \citenamefont {Igumenshchev} \emph {et~al.}}]{campbell2017laser}%
  \BibitemOpen
  \bibfield  {author} {\bibinfo {author} {\bibfnamefont {E.}~\bibnamefont
  {Campbell}}, \bibinfo {author} {\bibfnamefont {V.}~\bibnamefont {Goncharov}},
  \bibinfo {author} {\bibfnamefont {T.}~\bibnamefont {Sangster}}, \bibinfo
  {author} {\bibfnamefont {S.}~\bibnamefont {Regan}}, \bibinfo {author}
  {\bibfnamefont {P.}~\bibnamefont {Radha}}, \bibinfo {author} {\bibfnamefont
  {R.}~\bibnamefont {Betti}}, \bibinfo {author} {\bibfnamefont
  {J.}~\bibnamefont {Myatt}}, \bibinfo {author} {\bibfnamefont
  {D.}~\bibnamefont {Froula}}, \bibinfo {author} {\bibfnamefont
  {M.}~\bibnamefont {Rosenberg}}, \bibinfo {author} {\bibfnamefont
  {I.}~\bibnamefont {Igumenshchev}}, \emph {et~al.},\ }\bibfield  {title}
  {\bibinfo {title} {Laser-direct-drive program: Promise, challenge, and path
  forward},\ }\href@noop {} {\bibfield  {journal} {\bibinfo  {journal} {Matter
  and Radiation at Extremes}\ }\textbf {\bibinfo {volume} {2}},\ \bibinfo
  {pages} {37} (\bibinfo {year} {2017})}\BibitemShut {NoStop}%
\bibitem [{\citenamefont {Clark}\ \emph {et~al.}(2019)\citenamefont {Clark},
  \citenamefont {Weber}, \citenamefont {Milovich}, \citenamefont {Pak},
  \citenamefont {Casey}, \citenamefont {Hammel}, \citenamefont {Ho},
  \citenamefont {Jones}, \citenamefont {Koning}, \citenamefont {Kritcher} \emph
  {et~al.}}]{clark2019three}%
  \BibitemOpen
  \bibfield  {author} {\bibinfo {author} {\bibfnamefont {D.}~\bibnamefont
  {Clark}}, \bibinfo {author} {\bibfnamefont {C.}~\bibnamefont {Weber}},
  \bibinfo {author} {\bibfnamefont {J.}~\bibnamefont {Milovich}}, \bibinfo
  {author} {\bibfnamefont {A.}~\bibnamefont {Pak}}, \bibinfo {author}
  {\bibfnamefont {D.}~\bibnamefont {Casey}}, \bibinfo {author} {\bibfnamefont
  {B.}~\bibnamefont {Hammel}}, \bibinfo {author} {\bibfnamefont
  {D.}~\bibnamefont {Ho}}, \bibinfo {author} {\bibfnamefont {O.}~\bibnamefont
  {Jones}}, \bibinfo {author} {\bibfnamefont {J.}~\bibnamefont {Koning}},
  \bibinfo {author} {\bibfnamefont {A.}~\bibnamefont {Kritcher}}, \emph
  {et~al.},\ }\bibfield  {title} {\bibinfo {title} {Three-dimensional modeling
  and hydrodynamic scaling of national ignition facility implosions},\
  }\href@noop {} {\bibfield  {journal} {\bibinfo  {journal} {Physics of
  Plasmas}\ }\textbf {\bibinfo {volume} {26}},\ \bibinfo {pages} {050601}
  (\bibinfo {year} {2019})}\BibitemShut {NoStop}%
\bibitem [{\citenamefont {Craxton}\ \emph {et~al.}(2015)\citenamefont
  {Craxton}, \citenamefont {Anderson}, \citenamefont {Boehly}, \citenamefont
  {Goncharov}, \citenamefont {Harding}, \citenamefont {Knauer}, \citenamefont
  {McCrory}, \citenamefont {McKenty}, \citenamefont {Meyerhofer}, \citenamefont
  {Myatt} \emph {et~al.}}]{craxton2015direct}%
  \BibitemOpen
  \bibfield  {author} {\bibinfo {author} {\bibfnamefont {R.}~\bibnamefont
  {Craxton}}, \bibinfo {author} {\bibfnamefont {K.}~\bibnamefont {Anderson}},
  \bibinfo {author} {\bibfnamefont {T.}~\bibnamefont {Boehly}}, \bibinfo
  {author} {\bibfnamefont {V.}~\bibnamefont {Goncharov}}, \bibinfo {author}
  {\bibfnamefont {D.}~\bibnamefont {Harding}}, \bibinfo {author} {\bibfnamefont
  {J.}~\bibnamefont {Knauer}}, \bibinfo {author} {\bibfnamefont
  {R.}~\bibnamefont {McCrory}}, \bibinfo {author} {\bibfnamefont
  {P.}~\bibnamefont {McKenty}}, \bibinfo {author} {\bibfnamefont
  {D.}~\bibnamefont {Meyerhofer}}, \bibinfo {author} {\bibfnamefont
  {J.}~\bibnamefont {Myatt}}, \emph {et~al.},\ }\bibfield  {title} {\bibinfo
  {title} {Direct-drive inertial confinement fusion: A review},\ }\href@noop {}
  {\bibfield  {journal} {\bibinfo  {journal} {Physics of Plasmas}\ }\textbf
  {\bibinfo {volume} {22}},\ \bibinfo {pages} {110501} (\bibinfo {year}
  {2015})}\BibitemShut {NoStop}%
\bibitem [{\citenamefont {Hurricane}\ \emph {et~al.}(2019)\citenamefont
  {Hurricane}, \citenamefont {Springer}, \citenamefont {Patel}, \citenamefont
  {Callahan}, \citenamefont {Baker}, \citenamefont {Casey}, \citenamefont
  {Divol}, \citenamefont {D{\"o}ppner}, \citenamefont {Hinkel}, \citenamefont
  {Hohenberger} \emph {et~al.}}]{hurricane2019approaching}%
  \BibitemOpen
  \bibfield  {author} {\bibinfo {author} {\bibfnamefont {O.}~\bibnamefont
  {Hurricane}}, \bibinfo {author} {\bibfnamefont {P.}~\bibnamefont {Springer}},
  \bibinfo {author} {\bibfnamefont {P.}~\bibnamefont {Patel}}, \bibinfo
  {author} {\bibfnamefont {D.}~\bibnamefont {Callahan}}, \bibinfo {author}
  {\bibfnamefont {K.}~\bibnamefont {Baker}}, \bibinfo {author} {\bibfnamefont
  {D.}~\bibnamefont {Casey}}, \bibinfo {author} {\bibfnamefont
  {L.}~\bibnamefont {Divol}}, \bibinfo {author} {\bibfnamefont
  {T.}~\bibnamefont {D{\"o}ppner}}, \bibinfo {author} {\bibfnamefont
  {D.}~\bibnamefont {Hinkel}}, \bibinfo {author} {\bibfnamefont
  {M.}~\bibnamefont {Hohenberger}}, \emph {et~al.},\ }\bibfield  {title}
  {\bibinfo {title} {Approaching a burning plasma on the nif},\ }\href@noop {}
  {\bibfield  {journal} {\bibinfo  {journal} {Physics of Plasmas}\ }\textbf
  {\bibinfo {volume} {26}},\ \bibinfo {pages} {052704} (\bibinfo {year}
  {2019})}\BibitemShut {NoStop}%
\bibitem [{\citenamefont {Kline}\ \emph {et~al.}(2019)\citenamefont {Kline},
  \citenamefont {Batha}, \citenamefont {Benedetti}, \citenamefont {Bennett},
  \citenamefont {Bhandarkar}, \citenamefont {Hopkins}, \citenamefont {Biener},
  \citenamefont {Biener}, \citenamefont {Bionta}, \citenamefont {Bond} \emph
  {et~al.}}]{kline2019progress}%
  \BibitemOpen
  \bibfield  {author} {\bibinfo {author} {\bibfnamefont {J.}~\bibnamefont
  {Kline}}, \bibinfo {author} {\bibfnamefont {S.}~\bibnamefont {Batha}},
  \bibinfo {author} {\bibfnamefont {L.}~\bibnamefont {Benedetti}}, \bibinfo
  {author} {\bibfnamefont {D.}~\bibnamefont {Bennett}}, \bibinfo {author}
  {\bibfnamefont {S.}~\bibnamefont {Bhandarkar}}, \bibinfo {author}
  {\bibfnamefont {L.~B.}\ \bibnamefont {Hopkins}}, \bibinfo {author}
  {\bibfnamefont {J.}~\bibnamefont {Biener}}, \bibinfo {author} {\bibfnamefont
  {M.}~\bibnamefont {Biener}}, \bibinfo {author} {\bibfnamefont
  {R.}~\bibnamefont {Bionta}}, \bibinfo {author} {\bibfnamefont
  {E.}~\bibnamefont {Bond}}, \emph {et~al.},\ }\bibfield  {title} {\bibinfo
  {title} {Progress of indirect drive inertial confinement fusion in the united
  states},\ }\href@noop {} {\bibfield  {journal} {\bibinfo  {journal} {Nuclear
  Fusion}\ }\textbf {\bibinfo {volume} {59}},\ \bibinfo {pages} {112018}
  (\bibinfo {year} {2019})}\BibitemShut {NoStop}%
\bibitem [{\citenamefont {Lindl}(1995)}]{lindl1995development}%
  \BibitemOpen
  \bibfield  {author} {\bibinfo {author} {\bibfnamefont {J.}~\bibnamefont
  {Lindl}},\ }\bibfield  {title} {\bibinfo {title} {Development of the
  indirect-drive approach to inertial confinement fusion and the target physics
  basis for ignition and gain},\ }\href@noop {} {\bibfield  {journal} {\bibinfo
   {journal} {Physics of plasmas}\ }\textbf {\bibinfo {volume} {2}},\ \bibinfo
  {pages} {3933} (\bibinfo {year} {1995})}\BibitemShut {NoStop}%
\bibitem [{\citenamefont {Merritt}\ \emph {et~al.}(2019)\citenamefont
  {Merritt}, \citenamefont {Sauppe}, \citenamefont {Loomis}, \citenamefont
  {Cardenas}, \citenamefont {Montgomery}, \citenamefont {Daughton},
  \citenamefont {Wilson}, \citenamefont {Kline}, \citenamefont {Khan},
  \citenamefont {Schoff} \emph {et~al.}}]{merritt2019experimental}%
  \BibitemOpen
  \bibfield  {author} {\bibinfo {author} {\bibfnamefont {E.~C.}\ \bibnamefont
  {Merritt}}, \bibinfo {author} {\bibfnamefont {J.~P.}\ \bibnamefont {Sauppe}},
  \bibinfo {author} {\bibfnamefont {E.~N.}\ \bibnamefont {Loomis}}, \bibinfo
  {author} {\bibfnamefont {T.}~\bibnamefont {Cardenas}}, \bibinfo {author}
  {\bibfnamefont {D.~S.}\ \bibnamefont {Montgomery}}, \bibinfo {author}
  {\bibfnamefont {W.~S.}\ \bibnamefont {Daughton}}, \bibinfo {author}
  {\bibfnamefont {D.~C.}\ \bibnamefont {Wilson}}, \bibinfo {author}
  {\bibfnamefont {J.~L.}\ \bibnamefont {Kline}}, \bibinfo {author}
  {\bibfnamefont {S.~F.}\ \bibnamefont {Khan}}, \bibinfo {author}
  {\bibfnamefont {M.}~\bibnamefont {Schoff}}, \emph {et~al.},\ }\bibfield
  {title} {\bibinfo {title} {Experimental study of energy transfer in double
  shell implosions},\ }\href@noop {} {\bibfield  {journal} {\bibinfo  {journal}
  {Physics of Plasmas}\ }\textbf {\bibinfo {volume} {26}},\ \bibinfo {pages}
  {052702} (\bibinfo {year} {2019})}\BibitemShut {NoStop}%
\bibitem [{\citenamefont {Montgomery}(2016)}]{montgomery2016two}%
  \BibitemOpen
  \bibfield  {author} {\bibinfo {author} {\bibfnamefont {D.~S.}\ \bibnamefont
  {Montgomery}},\ }\bibfield  {title} {\bibinfo {title} {Two decades of
  progress in understanding and control of laser plasma instabilities in
  indirect drive inertial fusion},\ }\href@noop {} {\bibfield  {journal}
  {\bibinfo  {journal} {Physics of Plasmas}\ }\textbf {\bibinfo {volume}
  {23}},\ \bibinfo {pages} {055601} (\bibinfo {year} {2016})}\BibitemShut
  {NoStop}%
\bibitem [{\citenamefont {Myatt}\ \emph {et~al.}(2017)\citenamefont {Myatt},
  \citenamefont {Follett}, \citenamefont {Shaw}, \citenamefont {Edgell},
  \citenamefont {Froula}, \citenamefont {Igumenshchev},\ and\ \citenamefont
  {Goncharov}}]{myatt2017wave}%
  \BibitemOpen
  \bibfield  {author} {\bibinfo {author} {\bibfnamefont {J.}~\bibnamefont
  {Myatt}}, \bibinfo {author} {\bibfnamefont {R.}~\bibnamefont {Follett}},
  \bibinfo {author} {\bibfnamefont {J.}~\bibnamefont {Shaw}}, \bibinfo {author}
  {\bibfnamefont {D.}~\bibnamefont {Edgell}}, \bibinfo {author} {\bibfnamefont
  {D.}~\bibnamefont {Froula}}, \bibinfo {author} {\bibfnamefont
  {I.}~\bibnamefont {Igumenshchev}},\ and\ \bibinfo {author} {\bibfnamefont
  {V.}~\bibnamefont {Goncharov}},\ }\bibfield  {title} {\bibinfo {title} {A
  wave-based model for cross-beam energy transfer in direct-drive inertial
  confinement fusion},\ }\href@noop {} {\bibfield  {journal} {\bibinfo
  {journal} {Physics of Plasmas}\ }\textbf {\bibinfo {volume} {24}},\ \bibinfo
  {pages} {056308} (\bibinfo {year} {2017})}\BibitemShut {NoStop}%
\bibitem [{\citenamefont {Tabak}\ \emph {et~al.}(1994)\citenamefont {Tabak},
  \citenamefont {Hammer}, \citenamefont {Glinsky}, \citenamefont {Kruer},
  \citenamefont {Wilks}, \citenamefont {Woodworth}, \citenamefont {Campbell},
  \citenamefont {Perry},\ and\ \citenamefont {Mason}}]{tabak1994ignition}%
  \BibitemOpen
  \bibfield  {author} {\bibinfo {author} {\bibfnamefont {M.}~\bibnamefont
  {Tabak}}, \bibinfo {author} {\bibfnamefont {J.}~\bibnamefont {Hammer}},
  \bibinfo {author} {\bibfnamefont {M.~E.}\ \bibnamefont {Glinsky}}, \bibinfo
  {author} {\bibfnamefont {W.~L.}\ \bibnamefont {Kruer}}, \bibinfo {author}
  {\bibfnamefont {S.~C.}\ \bibnamefont {Wilks}}, \bibinfo {author}
  {\bibfnamefont {J.}~\bibnamefont {Woodworth}}, \bibinfo {author}
  {\bibfnamefont {E.~M.}\ \bibnamefont {Campbell}}, \bibinfo {author}
  {\bibfnamefont {M.~D.}\ \bibnamefont {Perry}},\ and\ \bibinfo {author}
  {\bibfnamefont {R.~J.}\ \bibnamefont {Mason}},\ }\bibfield  {title} {\bibinfo
  {title} {Ignition and high gain with ultrapowerful lasers},\ }\href@noop {}
  {\bibfield  {journal} {\bibinfo  {journal} {Physics of Plasmas}\ }\textbf
  {\bibinfo {volume} {1}},\ \bibinfo {pages} {1626} (\bibinfo {year}
  {1994})}\BibitemShut {NoStop}%
\bibitem [{\citenamefont {Vashistha}\ \emph {et~al.}(2020)\citenamefont
  {Vashistha}, \citenamefont {Mandal}, \citenamefont {Kumar}, \citenamefont
  {Shukla},\ and\ \citenamefont {Das}}]{vashistha2020new}%
  \BibitemOpen
  \bibfield  {author} {\bibinfo {author} {\bibfnamefont {A.}~\bibnamefont
  {Vashistha}}, \bibinfo {author} {\bibfnamefont {D.}~\bibnamefont {Mandal}},
  \bibinfo {author} {\bibfnamefont {A.}~\bibnamefont {Kumar}}, \bibinfo
  {author} {\bibfnamefont {C.}~\bibnamefont {Shukla}},\ and\ \bibinfo {author}
  {\bibfnamefont {A.}~\bibnamefont {Das}},\ }\bibfield  {title} {\bibinfo
  {title} {A new mechanism of direct coupling of laser energy to ions},\
  }\href@noop {} {\bibfield  {journal} {\bibinfo  {journal} {New Journal of
  Physics}\ }\textbf {\bibinfo {volume} {22}},\ \bibinfo {pages} {063023}
  (\bibinfo {year} {2020})}\BibitemShut {NoStop}%
\bibitem [{\citenamefont {Nakamura}\ \emph {et~al.}(2018)\citenamefont
  {Nakamura}, \citenamefont {Ikeda}, \citenamefont {Sawabe}, \citenamefont
  {Matsuda},\ and\ \citenamefont {Takeyama}}]{nakamura2018record}%
  \BibitemOpen
  \bibfield  {author} {\bibinfo {author} {\bibfnamefont {D.}~\bibnamefont
  {Nakamura}}, \bibinfo {author} {\bibfnamefont {A.}~\bibnamefont {Ikeda}},
  \bibinfo {author} {\bibfnamefont {H.}~\bibnamefont {Sawabe}}, \bibinfo
  {author} {\bibfnamefont {Y.}~\bibnamefont {Matsuda}},\ and\ \bibinfo {author}
  {\bibfnamefont {S.}~\bibnamefont {Takeyama}},\ }\bibfield  {title} {\bibinfo
  {title} {Record indoor magnetic field of 1200 t generated by electromagnetic
  flux-compression},\ }\href@noop {} {\bibfield  {journal} {\bibinfo  {journal}
  {Review of Scientific Instruments}\ }\textbf {\bibinfo {volume} {89}},\
  \bibinfo {pages} {095106} (\bibinfo {year} {2018})}\BibitemShut {NoStop}%
\bibitem [{\citenamefont {Korneev}\ \emph {et~al.}(2015)\citenamefont
  {Korneev}, \citenamefont {d'Humi{\`e}res},\ and\ \citenamefont
  {Tikhonchuk}}]{korneev2015gigagauss}%
  \BibitemOpen
  \bibfield  {author} {\bibinfo {author} {\bibfnamefont {P.}~\bibnamefont
  {Korneev}}, \bibinfo {author} {\bibfnamefont {E.}~\bibnamefont
  {d'Humi{\`e}res}},\ and\ \bibinfo {author} {\bibfnamefont {V.}~\bibnamefont
  {Tikhonchuk}},\ }\bibfield  {title} {\bibinfo {title} {Gigagauss-scale
  quasistatic magnetic field generation in a snail-shaped target},\ }\href@noop
  {} {\bibfield  {journal} {\bibinfo  {journal} {Physical Review E}\ }\textbf
  {\bibinfo {volume} {91}},\ \bibinfo {pages} {043107} (\bibinfo {year}
  {2015})}\BibitemShut {NoStop}%
\bibitem [{\citenamefont {Zosa}\ \emph {et~al.}(2022)\citenamefont {Zosa},
  \citenamefont {Gu},\ and\ \citenamefont {Murakami}}]{zosa2022100}%
  \BibitemOpen
  \bibfield  {author} {\bibinfo {author} {\bibfnamefont {M.-A.}\ \bibnamefont
  {Zosa}}, \bibinfo {author} {\bibfnamefont {Y.-J.}\ \bibnamefont {Gu}},\ and\
  \bibinfo {author} {\bibfnamefont {M.}~\bibnamefont {Murakami}},\ }\bibfield
  {title} {\bibinfo {title} {100-kt magnetic field generation using paisley
  targets by femtosecond laser--plasma interactions},\ }\href@noop {}
  {\bibfield  {journal} {\bibinfo  {journal} {Applied Physics Letters}\
  }\textbf {\bibinfo {volume} {120}},\ \bibinfo {pages} {132403} (\bibinfo
  {year} {2022})}\BibitemShut {NoStop}%
\bibitem [{\citenamefont {Mandal}\ \emph
  {et~al.}(2021{\natexlab{a}})\citenamefont {Mandal}, \citenamefont
  {Vashistha},\ and\ \citenamefont {Das}}]{mandal2021transparency}%
  \BibitemOpen
  \bibfield  {author} {\bibinfo {author} {\bibfnamefont {D.}~\bibnamefont
  {Mandal}}, \bibinfo {author} {\bibfnamefont {A.}~\bibnamefont {Vashistha}},\
  and\ \bibinfo {author} {\bibfnamefont {A.}~\bibnamefont {Das}},\ }\bibfield
  {title} {\bibinfo {title} {Electromagnetic wave transparency of x mode in
  strongly magnetized plasma},\ }\bibfield  {journal} {\bibinfo  {journal}
  {Scientific Reports}\ }\textbf {\bibinfo {volume} {11}},\ \href
  {https://doi.org/10.1038/s41598-021-94029-3} {10.1038/s41598-021-94029-3}
  (\bibinfo {year} {2021}{\natexlab{a}})\BibitemShut {NoStop}%
\bibitem [{\citenamefont {Goswami}\ \emph {et~al.}(2021)\citenamefont
  {Goswami}, \citenamefont {Maity}, \citenamefont {Mandal}, \citenamefont
  {Vashistha},\ and\ \citenamefont {Das}}]{goswami2021ponderomotive}%
  \BibitemOpen
  \bibfield  {author} {\bibinfo {author} {\bibfnamefont {L.~P.}\ \bibnamefont
  {Goswami}}, \bibinfo {author} {\bibfnamefont {S.}~\bibnamefont {Maity}},
  \bibinfo {author} {\bibfnamefont {D.}~\bibnamefont {Mandal}}, \bibinfo
  {author} {\bibfnamefont {A.}~\bibnamefont {Vashistha}},\ and\ \bibinfo
  {author} {\bibfnamefont {A.}~\bibnamefont {Das}},\ }\bibfield  {title}
  {\bibinfo {title} {Ponderomotive force driven mechanism for electrostatic
  wave excitation and energy absorption of electromagnetic waves in overdense
  magnetized plasma},\ }\href {https://doi.org/10.1088/1361-6587/ac206a}
  {\bibfield  {journal} {\bibinfo  {journal} {Plasma Physics and Controlled
  Fusion}\ }\textbf {\bibinfo {volume} {63}},\ \bibinfo {pages} {115003}
  (\bibinfo {year} {2021})}\BibitemShut {NoStop}%
\bibitem [{\citenamefont {Mandal}\ \emph
  {et~al.}(2021{\natexlab{b}})\citenamefont {Mandal}, \citenamefont
  {Vashistha}, \citenamefont {Goswami},\ and\ \citenamefont
  {Das}}]{mandal2021electromagnetic}%
  \BibitemOpen
  \bibfield  {author} {\bibinfo {author} {\bibfnamefont {D.}~\bibnamefont
  {Mandal}}, \bibinfo {author} {\bibfnamefont {A.}~\bibnamefont {Vashistha}},
  \bibinfo {author} {\bibfnamefont {L.}~\bibnamefont {Goswami}},\ and\ \bibinfo
  {author} {\bibfnamefont {A.}~\bibnamefont {Das}},\ }\bibfield  {title}
  {\bibinfo {title} {Electromagnetic transparency in strongly magnetized
  plasmas},\ }\href@noop {} {\bibfield  {journal} {\bibinfo  {journal}
  {Bulletin of the American Physical Society}\ }\textbf {\bibinfo {volume}
  {66}} (\bibinfo {year} {2021}{\natexlab{b}})}\BibitemShut {NoStop}%
\bibitem [{\citenamefont {Ritchie}(1994)}]{ritchie1994relativistic}%
  \BibitemOpen
  \bibfield  {author} {\bibinfo {author} {\bibfnamefont {B.}~\bibnamefont
  {Ritchie}},\ }\bibfield  {title} {\bibinfo {title} {Relativistic
  self-focusing and channel formation in laser-plasma interactions},\
  }\href@noop {} {\bibfield  {journal} {\bibinfo  {journal} {Physical Review
  E}\ }\textbf {\bibinfo {volume} {50}},\ \bibinfo {pages} {R687} (\bibinfo
  {year} {1994})}\BibitemShut {NoStop}%
\bibitem [{\citenamefont {Osman}\ \emph {et~al.}(1999)\citenamefont {Osman},
  \citenamefont {Castillo},\ and\ \citenamefont
  {Hora}}]{osman1999relativistic}%
  \BibitemOpen
  \bibfield  {author} {\bibinfo {author} {\bibfnamefont {F.}~\bibnamefont
  {Osman}}, \bibinfo {author} {\bibfnamefont {R.}~\bibnamefont {Castillo}},\
  and\ \bibinfo {author} {\bibfnamefont {H.}~\bibnamefont {Hora}},\ }\bibfield
  {title} {\bibinfo {title} {Relativistic and ponderomotive self-focusing at
  laser--plasma interaction},\ }\href@noop {} {\bibfield  {journal} {\bibinfo
  {journal} {Journal of plasma physics}\ }\textbf {\bibinfo {volume} {61}},\
  \bibinfo {pages} {263} (\bibinfo {year} {1999})}\BibitemShut {NoStop}%
\bibitem [{\citenamefont {Young}\ \emph {et~al.}(1989)\citenamefont {Young},
  \citenamefont {Baldis}, \citenamefont {Johnston}, \citenamefont {Kruer},\
  and\ \citenamefont {Estabrook}}]{young1989filamentation}%
  \BibitemOpen
  \bibfield  {author} {\bibinfo {author} {\bibfnamefont {P.}~\bibnamefont
  {Young}}, \bibinfo {author} {\bibfnamefont {H.}~\bibnamefont {Baldis}},
  \bibinfo {author} {\bibfnamefont {T.}~\bibnamefont {Johnston}}, \bibinfo
  {author} {\bibfnamefont {W.}~\bibnamefont {Kruer}},\ and\ \bibinfo {author}
  {\bibfnamefont {K.}~\bibnamefont {Estabrook}},\ }\bibfield  {title} {\bibinfo
  {title} {Filamentation and second-harmonic emission in laser-plasma
  interactions},\ }\href@noop {} {\bibfield  {journal} {\bibinfo  {journal}
  {Physical review letters}\ }\textbf {\bibinfo {volume} {63}},\ \bibinfo
  {pages} {2812} (\bibinfo {year} {1989})}\BibitemShut {NoStop}%
\bibitem [{\citenamefont {Liu}\ \emph {et~al.}(2019)\citenamefont {Liu},
  \citenamefont {Tripathi},\ and\ \citenamefont {Eliasson}}]{liu2019high}%
  \BibitemOpen
  \bibfield  {author} {\bibinfo {author} {\bibfnamefont {C.}~\bibnamefont
  {Liu}}, \bibinfo {author} {\bibfnamefont {V.}~\bibnamefont {Tripathi}},\ and\
  \bibinfo {author} {\bibfnamefont {B.}~\bibnamefont {Eliasson}},\ }\href@noop
  {} {\emph {\bibinfo {title} {High-power laser-plasma interaction}}}\
  (\bibinfo  {publisher} {Cambridge University Press},\ \bibinfo {year}
  {2019})\BibitemShut {NoStop}%
\bibitem [{\citenamefont {Boyd}\ \emph {et~al.}(2003)\citenamefont {Boyd},
  \citenamefont {Boyd},\ and\ \citenamefont {Sanderson}}]{boyd2003physics}%
  \BibitemOpen
  \bibfield  {author} {\bibinfo {author} {\bibfnamefont {T.}~\bibnamefont
  {Boyd}}, \bibinfo {author} {\bibfnamefont {T.}~\bibnamefont {Boyd}},\ and\
  \bibinfo {author} {\bibfnamefont {J.}~\bibnamefont {Sanderson}},\ }\href@noop
  {} {\emph {\bibinfo {title} {The physics of plasmas}}}\ (\bibinfo
  {publisher} {Cambridge University Press},\ \bibinfo {year}
  {2003})\BibitemShut {NoStop}%
\bibitem [{\citenamefont {Goswami}\ \emph {et~al.}(2022)\citenamefont
  {Goswami}, \citenamefont {Dhalia}, \citenamefont {Juneja}, \citenamefont
  {Maity}, \citenamefont {Das},\ and\ \citenamefont
  {Das}}]{goswami2022observations}%
  \BibitemOpen
  \bibfield  {author} {\bibinfo {author} {\bibfnamefont {L.~P.}\ \bibnamefont
  {Goswami}}, \bibinfo {author} {\bibfnamefont {T.}~\bibnamefont {Dhalia}},
  \bibinfo {author} {\bibfnamefont {R.}~\bibnamefont {Juneja}}, \bibinfo
  {author} {\bibfnamefont {S.}~\bibnamefont {Maity}}, \bibinfo {author}
  {\bibfnamefont {S.}~\bibnamefont {Das}},\ and\ \bibinfo {author}
  {\bibfnamefont {A.}~\bibnamefont {Das}},\ }\bibfield  {title} {\bibinfo
  {title} {Observations of brillouin scattering process in particle-in-cell
  simulations for laser pulse interacting with magnetized overdense plasma},\
  }\href@noop {} {\bibfield  {journal} {\bibinfo  {journal} {Physica Scripta}\
  }\textbf {\bibinfo {volume} {98}},\ \bibinfo {pages} {015602} (\bibinfo
  {year} {2022})}\BibitemShut {NoStop}%
\bibitem [{\citenamefont {Chen}\ \emph {et~al.}(1984)\citenamefont {Chen} \emph
  {et~al.}}]{chen1984introduction}%
  \BibitemOpen
  \bibfield  {author} {\bibinfo {author} {\bibfnamefont {F.~F.}\ \bibnamefont
  {Chen}} \emph {et~al.},\ }\href@noop {} {\emph {\bibinfo {title}
  {Introduction to plasma physics and controlled fusion}}},\ Vol.~\bibinfo
  {volume} {1}\ (\bibinfo  {publisher} {Springer},\ \bibinfo {year}
  {1984})\BibitemShut {NoStop}%
\bibitem [{\citenamefont {Hemker}(2000)}]{hemker2000particle}%
  \BibitemOpen
  \bibfield  {author} {\bibinfo {author} {\bibfnamefont {R.~G.}\ \bibnamefont
  {Hemker}},\ }\href@noop {} {\emph {\bibinfo {title} {Particle-in-cell
  modeling of plasma-based accelerators in two and three dimensions}}}\
  (\bibinfo  {publisher} {University of California, Los Angeles},\ \bibinfo
  {year} {2000})\BibitemShut {NoStop}%
\bibitem [{\citenamefont {Fonseca}\ \emph {et~al.}(2002)\citenamefont
  {Fonseca}, \citenamefont {Silva}, \citenamefont {Tsung}, \citenamefont
  {Decyk}, \citenamefont {Lu}, \citenamefont {Ren}, \citenamefont {Mori},
  \citenamefont {Deng}, \citenamefont {Lee}, \citenamefont {Katsouleas} \emph
  {et~al.}}]{fonseca2002osiris}%
  \BibitemOpen
  \bibfield  {author} {\bibinfo {author} {\bibfnamefont {R.~A.}\ \bibnamefont
  {Fonseca}}, \bibinfo {author} {\bibfnamefont {L.~O.}\ \bibnamefont {Silva}},
  \bibinfo {author} {\bibfnamefont {F.~S.}\ \bibnamefont {Tsung}}, \bibinfo
  {author} {\bibfnamefont {V.~K.}\ \bibnamefont {Decyk}}, \bibinfo {author}
  {\bibfnamefont {W.}~\bibnamefont {Lu}}, \bibinfo {author} {\bibfnamefont
  {C.}~\bibnamefont {Ren}}, \bibinfo {author} {\bibfnamefont {W.~B.}\
  \bibnamefont {Mori}}, \bibinfo {author} {\bibfnamefont {S.}~\bibnamefont
  {Deng}}, \bibinfo {author} {\bibfnamefont {S.}~\bibnamefont {Lee}}, \bibinfo
  {author} {\bibfnamefont {T.}~\bibnamefont {Katsouleas}}, \emph {et~al.},\
  }\bibfield  {title} {\bibinfo {title} {Osiris: A three-dimensional, fully
  relativistic particle in cell code for modeling plasma based accelerators},\
  }in\ \href@noop {} {\emph {\bibinfo {booktitle} {International Conference on
  Computational Science}}}\ (\bibinfo {organization} {Springer},\ \bibinfo
  {year} {2002})\ pp.\ \bibinfo {pages} {342--351}\BibitemShut {NoStop}%
\bibitem [{\citenamefont {Fonseca}\ \emph {et~al.}(2008)\citenamefont
  {Fonseca}, \citenamefont {Martins}, \citenamefont {Silva}, \citenamefont
  {Tonge}, \citenamefont {Tsung},\ and\ \citenamefont {Mori}}]{fonseca2008one}%
  \BibitemOpen
  \bibfield  {author} {\bibinfo {author} {\bibfnamefont {R.}~\bibnamefont
  {Fonseca}}, \bibinfo {author} {\bibfnamefont {S.}~\bibnamefont {Martins}},
  \bibinfo {author} {\bibfnamefont {L.}~\bibnamefont {Silva}}, \bibinfo
  {author} {\bibfnamefont {J.}~\bibnamefont {Tonge}}, \bibinfo {author}
  {\bibfnamefont {F.}~\bibnamefont {Tsung}},\ and\ \bibinfo {author}
  {\bibfnamefont {W.}~\bibnamefont {Mori}},\ }\bibfield  {title} {\bibinfo
  {title} {One-to-one direct modeling of experiments and astrophysical
  scenarios: pushing the envelope on kinetic plasma simulations},\ }\href@noop
  {} {\bibfield  {journal} {\bibinfo  {journal} {Plasma Physics and Controlled
  Fusion}\ }\textbf {\bibinfo {volume} {50}},\ \bibinfo {pages} {124034}
  (\bibinfo {year} {2008})}\BibitemShut {NoStop}%
\end{thebibliography}
%

\end{document}